\documentclass[twocolumn,aps,superscriptaddress]{revtex4-1}
\usepackage{palatino}
\usepackage[latin1]{inputenc}
\usepackage{epsf}
\usepackage{amsmath,amssymb}
\usepackage{latexsym}
\usepackage{calc}
\usepackage{color}
\usepackage{graphicx}
\usepackage{hyperref}

\newcommand{\ben}{\begin{equation}}
\newcommand{\een}{\end{equation}}
\newcommand{\bea}{\begin{eqnarray}}
\newcommand{\eea}{\end{eqnarray}}

\def\sss{\scriptscriptstyle\rm}

\def\1s{_{1,\sss S}}
\def\2s{_{2,\sss S}}

\def\br{{\bf r}}
\def\bR{{\bf R}}

\def\bA{{\bf A}}

\def\dulr{{\underline{\underline{\bf r}}}}
\def\dulR{{\underline{\underline{\bf R}}}}
\def\dulq{{\underline{\underline{q}}}}

\def\eps{{\epsilon}}
\def\wpol{_{\rm wpol}}
\def\GD{_{\rm GD}}

\begin{document}
\title{Case Studies of the Time-Dependent Potential Energy Surface for Dynamics in Cavities}
\author{Phillip Martinez}
\thanks{P. Martinez and B. Rosenzweig contributed equally to this work.}
\affiliation{Department of Physics and Astronomy, Hunter College of the City University of New York, 695 Park Avenue, New York, New York 10065, USA}
\author{Bart Rosenzweig}
\thanks{P. Martinez and B. Rosenzweig contributed equally to this work.}
\affiliation{Department of Physics and Astronomy, Hunter College of the City University of New York, 695 Park Avenue, New York, New York 10065, USA}
\affiliation{Department of Mathematics and Statistics, Hunter College of the City University of New York, 695 Park Avenue, New York, New York 10065, USA}
\author{Norah M. Hoffmann}
\thanks{Present address: Department of Chemistry, Columbia University, New York, New York 10027 USA}
\affiliation{Department of Physics, Rutgers University, Newark 07102, New Jersey USA}
\author{Lionel Lacombe}
\affiliation{Department of Physics, Rutgers University, Newark 07102, New Jersey USA}
\author{Neepa T. Maitra}
\affiliation{Department of Physics, Rutgers University, Newark 07102, New Jersey USA}
\email{neepa.maitra@rutgers.edu}
\date{\today}
\begin{abstract}
The exact time-dependent potential energy surface driving the nuclear dynamics was recently shown to be a useful tool to understand and interpret the coupling of nuclei, electrons, and photons, in cavity settings. 
Here we provide a detailed analysis of its structure for exactly-solvable systems that model two phenomena: cavity-induced suppression of proton-coupled electron transfer and its dependence on the initial state, and cavity-induced electronic excitation.
We demonstrate the inadequacy of simply using a weighted average of polaritonic surfaces to determine the dynamics. Such a weighted average misses a crucial term that redistributes energy between the nuclear and the polaritonic systems, and this term can in fact become a predominant term in determining the nuclear dynamics when several polaritonic surfaces are involved. 
Evolving an ensemble of classical trajectories on the exact potential energy surface reproduces the nuclear wavepacket quite accurately while evolving on the weighted polaritonic surface fails after a short period of time.
The implications and prospects for application of mixed quantum-classical methods based on this surface are discussed. 
\end{abstract}

\maketitle

\section{Introduction}
\label{sec:intro}
The interesting fact that simply confining a quantum system increases its coupling to vacuum fluctuations of the radiation field so much as to change the system's properties has been known since the early days of quantum mechanics.  A key realization  was that spontaneous emission rates of a two-level system can be increased by placing it in a resonant cavity due to the enhanced light-matter coupling strength~\cite{Purcell1946}. The recent burst of activity in cavity quantum electrodynamics is largely due to impressive experimental advances in manipulating matter at the meso- and nano-scales, that  have driven the investigations towards exploring the possibilities of new cavity-mediated phenomena. The playing field is rich when considering enhanced light-matter interaction on top of the already complex interactions of real matter, as opposed to the simple two-level-like systems that were considered earlier~\cite{RMDCY18,HO20, FGG18, RTFAR18, HWMB19, KBM16, SHCCV18,BWCAS18,XRDCYWYZ20}.

Polaritonic phenomena involve a complex interplay of photonic, electronic, and nuclear degrees of freedom, but often we are particularly interested in the dynamics of one particular subsystem. In polaritonic chemistry applications, it is typically the nuclear (or ionic) system we are interested in, for example in investigating changes in chemical reactivity when molecules are confined. 
The exact factorization (EF) approach~\cite{Hunter74,Hunter75,AMG10,AMG12, HARM18,LHM19} offers a natural context in which to study this problem since it provides a rigorous and unique definition of potentials acting on a subsystem that exactly incorporate the effects of coupling to the other subsystems. In particular, one obtains 
a time-dependent Schr\"odinger equation (TDSE) for the nuclear wavefunction $\chi(\dulR, t)$, in which the potentials contain the complete effect of coupling to the photonic and electronic systems. The full, exact wavefunction of the system has the form of a single correlated product, $\chi(\dulR,t)\Phi_\dulR(\dulr,\dulq, t)$, where $\dulr,\dulq$ represent electronic and photonic coordinates respectively. This is quite in contrast to another exact representation for the full wavefunction, the Born-Huang expansion generalized to include photon degrees of freedom, which contains an infinite number of terms of correlated products, $\sum_j\chi^j(\dulR,t)\Phi_{\dulR}^j(\dulr,q)$; here the $\Phi_{\dulR}^j(\dulr,q)$ are the polaritonic eigenstates. Instead of the single TDSE for the nuclear system that appears in EF, one has a set of coupled equations to solve, and it is not possible to identify a potential that drives the nuclear motion. The concept of a single equation for the nuclear wavefunction, as arises in EF, is especially useful when turning to mixed quantum-classical methods where being able to unambiguously define a force  acting on a classical nuclear trajectory at each point is desirable.

The EF was originally derived with the static coupled electron-nuclear problem in mind~\cite{Hunter74,Hunter75,GG14}, while the time-dependent problem was introduced later~\cite{AMG10,AMG12}. Extensions of the EF idea to a variety of different settings have been made: e.g.  the static purely electronic problem~\cite{Hunter86} with connections to density functional theory~\cite{BBS89}, the time-dependent electronic problem with connections to the single-active electron approach~\cite{SG17}, and recently electronic Fock space to derive a new quantum embedding method~\cite{GZR18,LM20,RG19}. A reverse factorization for electron-nuclear problems has also been explored where the nuclear wavefunction is conditionally dependent on the electronic coordinate, which has an advantage when one is most interested in the electronic system~\cite{SAMYG14}. 

Refs.~\cite{HARM18,LHM19,AKT18} extended the EF approach to polaritonic systems, focussing on phenomena which are either primarily interesting for the photonic system~\cite{HARM18}, the electronic system~\cite{AKT18}, or the nuclear system~\cite{LHM19}. The present work expands on the latter. Ref.~\cite{LHM19} studied  the exact potential driving a model of cavity-induced suppression of proton coupled electron-transfer. The model molecule was photo-excited into the first excited Born-Oppenheimer state, and, when outside the cavity, the nuclear wavepacket evolves towards a narrow avoided crossing where there is some transfer of population to the ground-electronic state. Since in this state the electronic dipole is localized on the opposite side to where it was localized initially, the process is known as proton-coupled electron transfer.  
The enhanced spontaneous emission rate from the excited molecule when placed in an optical cavity led to a significant partial photon emission when the cavity resonance was tuned to the initial excitation frequency. The interplay between the partial photon emission, the partial electronic de-excitation, and coupling to nuclear motion led to less of the nuclear wavepacket evolving to the narrow avoided crossing, and hence a suppression of the proton-coupled electron-transfer. 
While the shape of the polaritonic surfaces suggest that such a phenomenon could occur, they alone could not predict the suppression, as  a weaker light-matter coupling strength led to practically identical surfaces but with  no suppression dynamics. What would be needed to determine the dynamics would be coupling matrix elements between the surfaces. On the other hand, the exact potential arising from the EF approach could distinguish clearly the suppression phenomenon.

In this paper, we present two further examples of the EF approach applied to polaritonic chemistry. In the first, we 
revisit the same model of Ref.~\cite{LHM19} but begin now in an excited polaritonic state instead of an electronically excited state. The dependence of cavity-altered phenomena on how the initial state of polaritonic system has not been widely explored. In particular, when there is an initial excitation in the system, a question is whether this excitation is done in the presence of the cavity such that the initial excitation is a polaritonic one, or whether it is done effectively outside the cavity and then the molecule is inserted into the cavity, in which case the initial excitation is a purely electronic one. 
 For our model of proton-coupled electron-transfer, we find suppression is observed when beginning in a polaritonic state, although not as much as for the initial purely electronically excited state, and we compare the exact potential from EF in the two cases and with the cavity-free case. 
The second example we consider is one of cavity-induced electronic excitation, where a photo-excited molecule is placed in a cavity whose fundamental frequency is resonant with a molecular frequency at two different geometries: at one nuclear configuration the mode frequency coincides with the ground and first electronic states while at the other nuclear configuration it coincides with the ground and second. Thus as an excited nuclear wavepacket on the first electronic state evolves through the first nuclear configuration into the second, an electronic excitation into the second excited Born-Oppenheimer state occurs, which is absent in the cavity-free case. With an eye to developing mixed quantum-classical approximations, we plot the force on the proton at different time snap-shots due to different components of this exact surface, especially since somewhat localized large step features appear in one of the components which may exaggerate its effect on the dynamics. A quasiclassical Wigner propagation of nuclear trajectories shows that dynamics on the exact potential reproduces the exact quantum dynamics accurately. 
We begin in Sec.~\ref{sec:EFA} with a brief review of the EF approach.

\section{Exact Factorization Approach}
\label{sec:EFA}
The essential statement of the exact factorization approach is very general: the complete wavefunction for  a system of coupled quantum subsystems can be factorized into a marginal wavefunction depending on one (or some) of the subsystem coordinates, and a conditional wavefunction of the other coordinates, parametrically dependent on the marginal degrees of freedom. 
The equations for the two parts depend on the choice of the marginal and conditional and how the Hamiltonian depends on them. 
For example, consider a system of interacting electrons only. In Ref.~\cite{SG17}, the factorization is done in real-space with a chosen number of electronic coordinates associated with the marginal, and the resulting equation for the marginal has a TDSE form. This is not the case in Ref.~\cite{LM20} where instead the factorization is done in Fock space with a chosen set of orbitals included in the marginal; instead, there an effective embedding Hamiltonian matrix is deduced  which is of interest to problems in strong correlation. 

Consider now any situation for which the components of the Hamiltonian associated with the chosen marginal degrees of freedom have the form of a kinetic term, $\hat{T} = -\sum_I\nabla_I^2/(2 M_I)$, plus potential terms that are multiplicative operators in the marginal coordinate. Then the EF equations take the same form as in the original electron-nuclear case~\cite{AMG10,AMG12}.  In particular, the equation for the marginal factor is a Schr\"odinger equation, containing scalar and vector potentials that incorporate exactly the effect of couplings to the other subsystems. 
Indeed, this is the case for a molecule in a cavity, when we take the non-relativistic photon-matter Hamiltonian in the dipole approximation in the length gauge~\cite{RTFAR18,RFPATR14,T13,MVH20}. 
The full Hamiltonian has the form
\ben
\hat{H} = \hat{T_n} +  \hat{H}_{\rm BO} + \hat{H}_p + \hat{V}_{pm} +\hat{V}_{\rm SP}\,,
\een
where $\hat{T_n}  = -\sum_I\nabla_I^2/(2 M_I)$ is the nuclear kinetic energy, 
$\hat{H}_{\rm BO}$ is the Born-Oppenheimer Hamiltonian of the molecule, and the other terms are arise from the photonic degrees of freedom. Note that we use atomic units ($e^2 = \hbar = m_e  = 1$) throughout this paper.
$\hat{H}_p$ is the ``free" photonic Hamiltonian, which, has the harmonic form
\ben
\hat{H}_p= \sum_\alpha\frac{1}{2}(\hat{p}^2_\alpha+ \omega^2_\alpha {\hat{q}}^2_\alpha)\,.
\een
Here the sum goes over all the modes $\alpha$ in the cavity, but in our examples we consider only a single mode of frequency $\omega_\alpha = \omega_c$, resonant with an electronic excitation frequency at a particular molecular geometry; $\hat{q}_\alpha$ is the displacement field coordinate, and $\hat{p}_\alpha = -i \partial_\alpha$. Recent work has shown that the dynamics can change significantly as more cavity modes are accounted for~\cite{HLRM20} but here we assume that the harmonics of this fundamental mode are high enough that they do not couple strongly with the molecule.   
The matter-photon coupling is represented by the term 
\ben
\hat{V}_{\rm pm}=\sum_\alpha^{2N_p}\omega_\alpha \hat{q}_\alpha {\vec\lambda_\alpha} \cdot \left(\sum_I^{N_n}Z_I\hat{\bR}_I - \sum_i^{N_e}\hat{\br}_i\right)\,,
\een
where $N_p$ is the number of photon modes, $N_{n,e}$ the number of nuclei and electrons respectively, and 
$\vec{\lambda_\alpha}$  the matter-photon coupling vector. Finally the self-polarization term is 
\ben
\hat{V}_{\rm SP} = \frac{1}{2}\sum_\alpha^{2N_p} \left( {\vec\lambda_\alpha} \cdot \left(\sum_I^{N_n}Z_I\hat{\bR}_I - \sum_i^{N_e}\hat{\br}_i\right)\right)^2\,.
\een
This has a negligible effect in our examples, due to the small value of $\lambda_\alpha$ that we use, so we will later safely neglect it. We note that the importance of this term grows as more modes are considered~\cite{HLRM20}, and its necessity in obtaining a consistent ground-state and maintaining gauge-invariance has been pointed out~\cite{SRRR20,RWRR18}. It should be borne in mind that the Hamiltonian above completely neglects any losses at the cavity boundaries, so the phenomena that we will observe will in reality be dampened somewhat. 

In the following we let $\dulR$ represent all nuclear coordinates (three degrees of freedom for each nucleus), $\dulr$ represent all electronic coordinates, and $\dulq$ all the displacement field coordinates for the photons (one degree of freedom for each mode). 
Taking the marginal as the nuclear degrees of freedom, 
\ben
\Psi(\dulr,\dulR,\dulq,t) = \chi(\dulR,t)\Phi_{\dulR}(\dulr,\dulq,t)
\label{eq:Psi_EF}
\een
yields~\cite{HARM18,LHM19,AMG10,AMG12,ACEJ13,AMG13}
\ben
\begin{aligned}[b]
\label{eq:Phi}
\left(\hat{H}_ {\rm BO} + \hat{H}_p +\hat{V}_{pm} + \hat{V}_{\rm SP} + \hat{U}_{\rm ep-n} - \eps(\dulR,t)\right)\Phi_\dulR(\dulr,\dulq,t) \\= i \partial_t  \Phi_\dulR(\dulr,\dulq,t)
\end{aligned}
\een
and
\ben
\begin{aligned}[b]
\left(\sum_{I=1}^{N_n}( -i\nabla_I + \bA_I(\dulR,t))^2/2M_I + \eps(\dulR,t)\right)\chi(\dulR,t) \\= i\partial_t\chi(\dulR,t)
\label{eq:chi}
\end{aligned}
\een
with the following coupling terms:
\newline
(i) the coupling operator between the electron-photon system and the nuclei
\bea
\nonumber
\hat{U}_{ep-n} =\sum_{I=1}^{N_n}\frac{1}{M_J}\left( \frac{(-i\nabla_I - \bA_I(\dulR,t))^2}{2}\right. + \\
\left.\left(\frac{-i\nabla_I\chi(\dulR,t)}{\chi(\dulR,t)} + \bA_I(\dulR,t)\right)\cdot\left(-i\nabla_I - \bA_I(\dulR,t)\right)\right)\,,
\label{eq:Uepn}
\eea
(ii) the time-dependent potential energy surface (TDPES)
\ben
\eps(\dulR,t) =\eps_{\rm wpol}(\bR,t) + \eps_{\rm kin}(\bR,t) + \eps_{\rm GD}(\bR, t) 
\label{eq:tdpes}
\een
which has the three components: the (self-polarization-included) weighted-polaritonic component,
\ben
\eps_{\rm wpol}(\bR,t) =\langle\Phi_\dulR\vert\hat{H}_ {\rm pol}+ \hat{V}_{\rm SP} \vert\Phi_\dulR\rangle_{\dulr,\dulq}\,,
\label{eq:wpol}
\een
where 
\ben
\hat{H}_{\rm pol} = \hat{H}_ {\rm BO} + \hat{H}_p +\hat{V}_{pm} 
\een
the kinetic-like term
\ben
\eps_{\rm kin} = \sum_{I=1}^{N_n}\frac{1}{2M_I}\langle\Phi_\dulR\vert\left(-\nabla_I - \bA_I(\dulR,t)\right)^2\vert\Phi_\dulR\rangle_{\dulr,\dulq}\,,
\label{eq:kin}
\een
and the gauge-dependent term
\ben
\eps_{\rm GD}(\dulR,t)=\langle\Phi_\dulR\vert-i\partial_t\vert\Phi_\dulR\rangle_{\dulr,\dulq}\,,
\label{eq:GD}
\een
and 
\newline
(iii) the time-dependent vector potential
\ben
\bA_I(\dulR,t) = \langle \Phi_\dulR\vert -i\nabla_I \Phi_\dulR\rangle_{\dulr,\dulq}
\label{eq:A}
\een
 Since the marginal factor $\chi(\dulR,t)$ in  Eq.~(\ref{eq:Psi_EF}) can be multiplied by a phase $e^{i\theta(\dulR,t)}$ while the conditional factor $\Phi_\dulR(\dulr,\dulq,t)$ is multiplied by the inverse phase $e^{-i\theta(\dulR,t)}$, without changing the product $\chi\Phi_\dulR$, there is a gauge-like dependence  to these equations~\cite{AMG10,AMG12}. One can choose different gauges in which $\bA_I(\dulR,t)$ is changed by the addition of $\nabla_I \theta(\dulR,t)$ while $\eps_{\rm GD}(\dulR,t)$ is changed by the addition of $\partial_t \theta(\dulR,t)$. 
The relation of the coupling terms that appear in the EF to the couplings that appear in a Born-Huang expansion may be found in Ref.~\cite{AMG10} for the electron-nuclear case.

We now note three simplifications to these general equations that are relevant for the case studies we will consider. 
First, our models here have only one nuclear degree of freedom, which means that we can always choose a gauge in which the vector potential $\bA(\dulR,t) = A(R,t) = 0$. 
With this gauge-choice, then the exact nuclear wavefunction satisfies a Schr\"odinger equation Eq.~(\ref{eq:chi}) with a purely scalar potential of Eq.~(\ref{eq:tdpes}), composed of the three components, Eq.~(\ref{eq:wpol})--Eq.~(\ref{eq:GD}).
Second, the kinetic component $\eps_{\rm kin}(R,t)$ is negligible compared to the other two terms, due to the factor of $1/M$ where $M$ is the nuclear mass. Our models consider a proton of mass 1836 times the mass of an electron.  This is in contrast to the cases where the electronic~\cite{SAMYG14,KAM15} or photonic~\cite{HARM18} systems are  chosen to be the marginal, where it certainly cannot be neglected. 
Third, we note that the self-polarization contribution to $\eps_{\rm wpol}(R,t)$ is negligible for the single-mode cases  and coupling strengths we are considering here; this term becomes important on the other hand when many cavity-modes are considered~\cite{HLRM20}.
Thus, the nuclear dynamics is driven by the Schr\"odinger equation~(\ref{eq:chi}) taking $A(R,t) = 0$ and $\eps(R,t) = \eps\wpol(R,t) + \eps\GD(R,t)$. 

\subsection{Interpreting the TDPES}
The TDPES is composed then of two potentials $\eps\wpol(R,t)$ and $\eps\GD(R,t)$ that completely govern the nuclear motion. Their dependence on the conditional electronic-photonic wavefunction $\Phi_R(\dulr,\dulq,t)$ embodies the correlation between the nuclear and electronic-photonic systems. It is instructive to expand $\Phi_R(\dulr,\dulq,t)$ in terms of the polaritonic eigenstates:
\ben
\Phi_R(\dulr,\dulq,t) =  \sum_k {\cal{C}}_k(R,t)\Phi^k_R(\dulr,\dulq)
\label{eq:cond-exp}
\een
where 
\ben
\hat{H}_{\rm pol}\Phi^k_R(\dulr,\dulq) = \eps_{{\rm pol},k}(R)\Phi^k_R(\dulr,\dulq) 
\label{eq:polstate}
\een
defines the polaritonic surfaces $\eps_{{\rm pol},k}(\bR)$ and polaritonic eigenstates $\Phi^k_R(\dulr,\dulq)$~\cite{GGF15,GGF16}. Then, due to orthonormality of the polaritonic eigenstates, one can decompose $\eps\wpol$ in terms of the polaritonic projections:
\ben
\eps\wpol(R,t) = \sum_k \vert {\cal{C}}_k(R,t) \vert^2 \eps_{{\rm pol},k}(R)
\een
The physical meaning of $\eps\wpol$ is then clear as a weighted average of the polaritonic surfaces, with the weights given by the projections of the conditional electronic-photonic wavefunction on the polaritonic surfaces. With mixed quantum-classical methods in mind, this is somewhat reminiscent of the form of the effective potential the classical nuclei experience in Ehrenfest dynamics, however with the crucial difference in that the coefficients are here {\it spatially-dependent}. Terms in the electronic equation yield gradients  of these terms (with respect to the nuclear coordinate) which allow different trajectories making up the nuclear wavepacket in a multi-trajectory approach to go in different directions. This results in the possibility of wavepacket splitting when there is partial occupation of different electronic surfaces, which does not happen in a multi-trajectory Ehrenfest approach~\cite{AMAG16,AASMMG15,GAM18,AC19}. As in Ref.~\cite{AASMMG15}, we will see in Secs.~\ref{sec:pcet} and~\ref{sec:elex} that although wavepacket splitting  occurs to some extent when $\eps\wpol$ is used to propagate,  $\eps\GD$ greatly increases this effect and is needed to get accurate  dynamics.

Similarly, expanding the gauge-dependent term in terms of polaritonic components, 

\begin{equation}
\eps\GD(R,t) = -\sum_k {\cal{C}}_k^*(R,t) i\partial_t{\cal{C}}_k(R,t).
\end{equation}

The meaning and role of $\eps\GD$ is less directly clear from its definition. Instead, some insight is gained from comparing the total energy of the system $\langle \hat{H} \rangle$ with the "marginal energy" defined via $\langle\hat{H}_{\rm nuc}\rangle$ where $H_{\rm nuc}$ is the Hamiltonian in the nuclear Schr\"odinger equation, governing the motion of the nuclear wavepacket.
We have
\bea
\langle\hat{H}\rangle &=& \langle\hat{H}_{\rm pol}\rangle  + \langle\hat{T}_n\rangle \\
\nonumber
&=& \int \vert\chi\vert^2\eps\wpol \,dR + \langle\hat{T}^{\rm marg}_n\rangle + \int \vert\chi\vert^2 \langle \hat{U}_{ep-n}\rangle_{\dulr,\dulq}\,dR
\label{eq:energy}
\eea
where $\hat{T}_n^{\rm marg} = -\int \chi^*(R,t)\frac{\nabla^2}{2M}\chi(R,t) dR$ is the kinetic energy of the marginal system, that differs from the nuclear kinetic energy $ \langle\hat{T}_n\rangle =-\int d\dulr d\dulq \Psi^*(\dulr,R,\dulq,t) \frac{\nabla_R^2}{2M} \Psi(\dulr,R,\dulq,t) $ by the last term in Eq.~(\ref{eq:energy})~\cite{AMG12}. The expectation value of $\hat{U}_{ep-n}$ however is very small compared with the other terms when the marginal is chosen as the nuclear system, due to the $1/M$ factor in Eq.~(\ref{eq:Uepn})~\cite{AMAG16}. (Note that only the first term of Eq.~(\ref{eq:Uepn}) contributes to its expectation value). 
This means, that the total energy is essentially
\ben
\langle\hat{H}\rangle =\int \vert\chi(R,t)\vert^2\eps\wpol (R,t)\,dR + \langle\hat{T}^{\rm marg}_n\rangle
\een
On the other hand, the ``marginal energy", $\langle \hat{H}_{\rm nuc}\rangle$, defined as that associated with the Hamiltonian driving the marginal motion, 
\bea
\langle \hat{H}_{\rm nuc}\rangle &=& \int \vert\chi(R,t)\vert^2\eps(R,t) \,dR + \langle\hat{T}^{\rm marg}_n\rangle\\
&=&\langle\hat{H}\rangle + \int \vert\chi(R,t)\vert^2\eps\GD (R,t)\,dR 
\eea
That is, $\eps\GD(R,t)$ does not contribute to the total energy but instead redistributes energy between the electron-photon system and the nuclear system.

To get a qualitative sense of its structure, we allow ourselves to extend the ideas above to the energy density rather than the total integrated energy. 
We distinguish two cases. 
First, when the polaritonic character is distinct in a piecewise localized way, i.e. one polaritonic character dominates the wavefunction in one region of space, while another dominates it in a neighboring region. For definiteness, let us say the polaritonic energy is locally larger on the left than on the right.  Then we expect $\langle\hat{T}^{\rm marg}_n\rangle$ would be smaller on the left than on the right, and this is enabled by a step up in $\eps\GD$ from left to right in the transition region. 
This has been observed in the electron-nuclear case~\cite{AASG13,AASMMG15}, and we will observe it again here in our polaritonic examples in the next section. 
In the second case, where there is a mixed polaritonic character, the expected structure of $\eps\GD(R,t)$ is generally less clear. Here different parts of the nuclear wavepacket are associated with different polaritonic surfaces in the same region of space. 
In the cases where the mean-field nature of $\eps\wpol(R,t)$ ``washes out" distinct structures, we have found $\eps\GD(R,t)$ can be of paramount importance. For example, when part of the nuclear wavepacket associated with a polaritonic surface begins to reflect from this surface, while the other parts do not,  a dynamical step appears in $\eps\GD(R,t)$ to enable the reflection of the part. Because of the average nature of $\eps\wpol(R,t)$ the different curvatures of different polaritonic surfaces ``under" the nuclear wavepacket are not always evident.

The equations of the exact factorization approach are no easier to solve than the original full Schr\"odinger equation; on the contrary, practical numerical schemes to solve them are currently elusive~\cite{GLM19,Lorin20}.  Rather, the practical power of the approach is due to providing rigorous definitions of coupling potentials, which then offer a clear and well-founded starting point for approximation schemes, as in Refs.~\cite{MAG15,AMAG16,MATG17,HLM18, FMK19,FPMC19,FPMK18}. 
Studies of the exact potentials in cases where they could be found~\cite{AASG13,AASMMG15}  were very instructive for these developments. Like in those works, we here obtain the exact potentials for the polaritonic case by inversion. 
That is, we first compute the full matter-photon wavefunction, which is possible for the systems of one nuclear, one electronic, and one photonic degree of freedom that we study. We then obtain 
the nuclear wavefunction $\chi(R,t)=\vert\chi(R,t)\vert e^{iS(R,t)}$,
which corresponds to the square root of the nuclear density times a phase. The nuclear density is obtained by integration, $\vert\chi(R,t)\vert^2 = \int dr dq \vert \Psi(r,R,q,t)\vert^2$. 
The phase is  fixed by the gauge choice, which for $A(R,t)=0$
gives $S(R,t)=\int^{R}dR'\;\Im(\langle\Psi\vert\partial_{R}\vert\Psi\rangle)/\vert\chi(R,t)\vert^{2}$~\cite{AASMMG15}. 
Once $\chi(R,t)$ is obtained, the rest is straightforward as $\Phi_{R}(r,q,t)=\Psi(r,R,q,t)/\chi(R,t)$
and we simply use Eqs.~(\ref{eq:tdpes})--(\ref{eq:GD}) with a simple two-point
stencil finite difference in time to compute the gauge-dependent part
of the potential.

\section{Model Hamiltonian}
\label{sec:model}
The Hamiltonian for the matter part of our model systems has the Shin-Metiu~\cite{SM95,FH97,FH97b} form:
\bea
\hat{H}_m = \hat{T}_n + \hat{H}_{\rm BO} = \hat{T}_n+\hat{T}_e+\hat{V}_m
\eea
where $\hat{T}_n=-\frac{1}{2M}\frac{\partial^2}{\partial R^2}$, $\hat{T}_e=-\frac{1}{2}\frac{\partial^2}{\partial r^2}$, and
\bea
\hat{V}_m = \sum_{\sigma = \pm 1}\Bigg(\frac{1}{|R+\frac{\sigma L}{2}|}-\frac{\rm erf\big(\frac{|r+\frac{\sigma L}{2}|}{a_\sigma}\big)}{|r+\frac{\sigma L}{2}|}\Bigg) - \frac{\rm erf\big(\frac{|R-r|}{a_f}\big)}{|R-r|}
\eea
which describes one electron and one proton moving along the axis between two fixed heavy ions separated by a distance $L$.

\begin{figure}
\includegraphics[width=1.0\columnwidth]{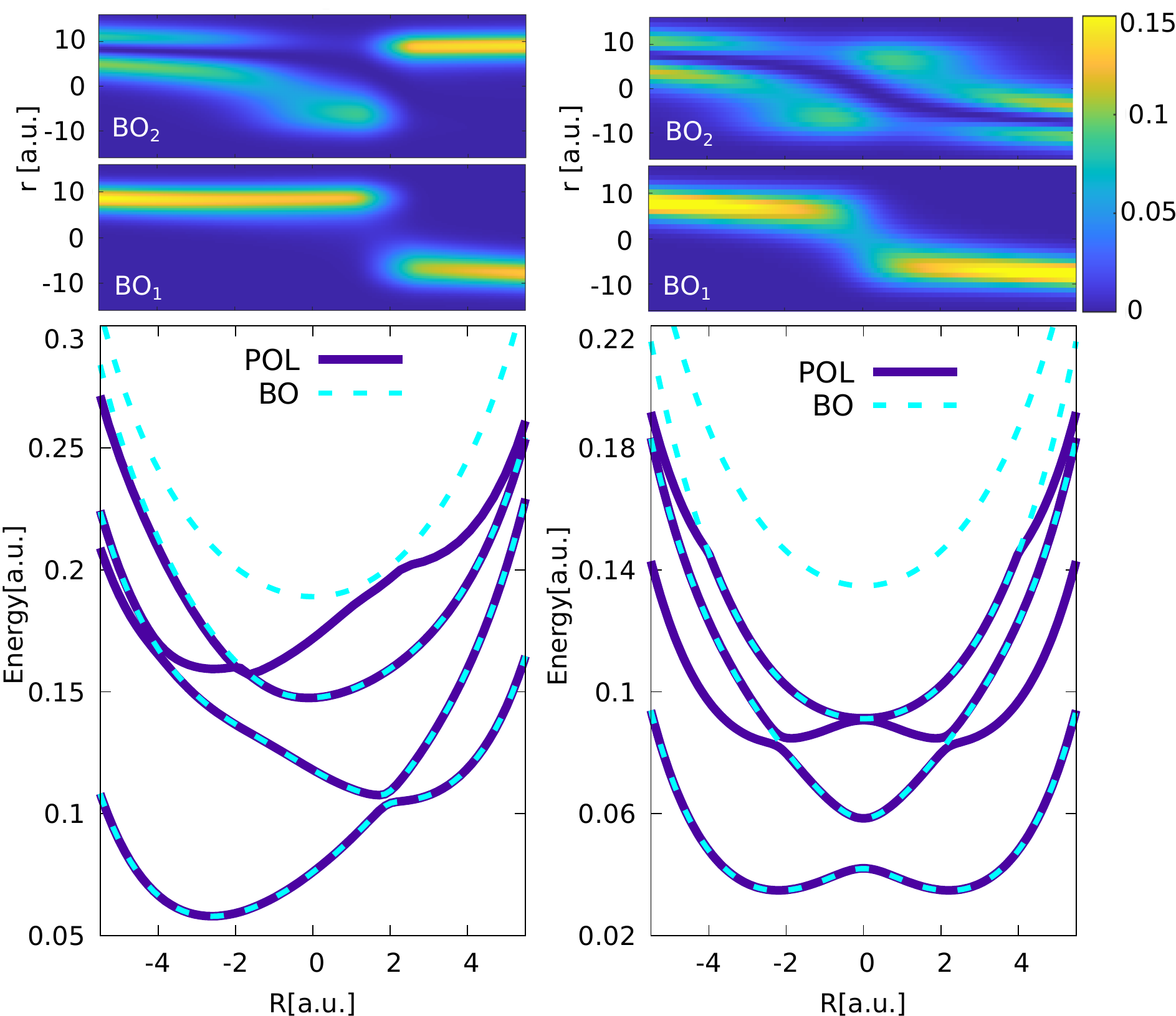}
\caption{Lower panels: BO (dashed) and polaritonic (solid) surfaces: cavity-induced suppression of PCET model (left) and cavity-induced electronic transition (right). The upper panels show the ground and first excited electronic BO wavefunctions, $\Phi^{{\rm BO},i}_{R}, i = 1,2$, for each case.}
\label{fig:bopol}
\end{figure}

We consider two sets of parameters, which will be used to illustrate two effects. 
The first effect (Sec.~\ref{sec:pcet}) is cavity-induced suppression of proton-coupled electron-transfer (PCET), and the parameters of the model are $L = 19.0$ a.u., $a_+ = 3.1$ a.u., $a_- = 4.0$ a.u., $a_f=5.0$ a.u., and $M=1836$ a.u., the proton mass, which was the set of parameters chosen in Ref.~\cite{LHM19}. The cavity frequency is chosen as $\omega_c= 0.1$a.u. and the photon-matter coupling strength as  $\lambda = 0.005$a.u. The second effect (Sec.~\ref{sec:elex}) is cavity-induced electronic excitation (ELEX), and the parameters chosen here are $L = 19.0$ a.u., $a_+ = 4.0$ a.u., $a_- = 4.0$ a.u., $a_f=5.0$ a.u., and $M=1836$ a.u, with the cavity frequency $\omega_c = 0.049$a.u. and again $\lambda = 0.005$a.u. 
These parameters are in a reasonable ballpark for plasmic nano gap cavities.
The first few BO and polaritonic surfaces are shown in Fig.~\ref{fig:bopol}. In all our examples the initial nuclear wavepacket is a Gaussian centered at $-4$a.u.

\subsection{Cavity-induced suppression of PCET}
\label{sec:pcet}
Ref.~\cite{LHM19} showed that when the Shin-Metiu molecule is placed in a cavity resonant with the initial BO energy difference, the PCET process is suppressed. Part of the nuclear wavepacket never makes it to the electron-nuclear avoided crossing region as it is associated with the emission of a photon. This behavior of the nuclear wavepacket was shown to be directly correlated with the curvature of the exact TDPES as it evolves in time, and impossible to predict from the shape of the static polaritonic surfaces alone. 
In Ref.~\cite{LHM19} the initial state was a Gaussian nuclear wavepacket placed on the second BO surface with no photon in the cavity, which would be the case if the excited molecule was prepared outside the cavity and then placed inside it.  Instead, if the molecule is placed inside the cavity before the excitation is done, beginning in a polaritonic state would be more appropriate than the purely electronically excited one. This is the case we consider in the present paper.  That is, the initial state has the form $\Psi(r,R,q,0) = {\cal{N}}e^{-2.85(R+4)^2}\Phi_R^{k=1}(r,q)$, where $\Phi_R^{k=1}(r,q)$ is the first excited polaritonic state (see Eq.~\ref{eq:polstate}). 

\begin{figure}
\includegraphics[width=1.0\columnwidth]{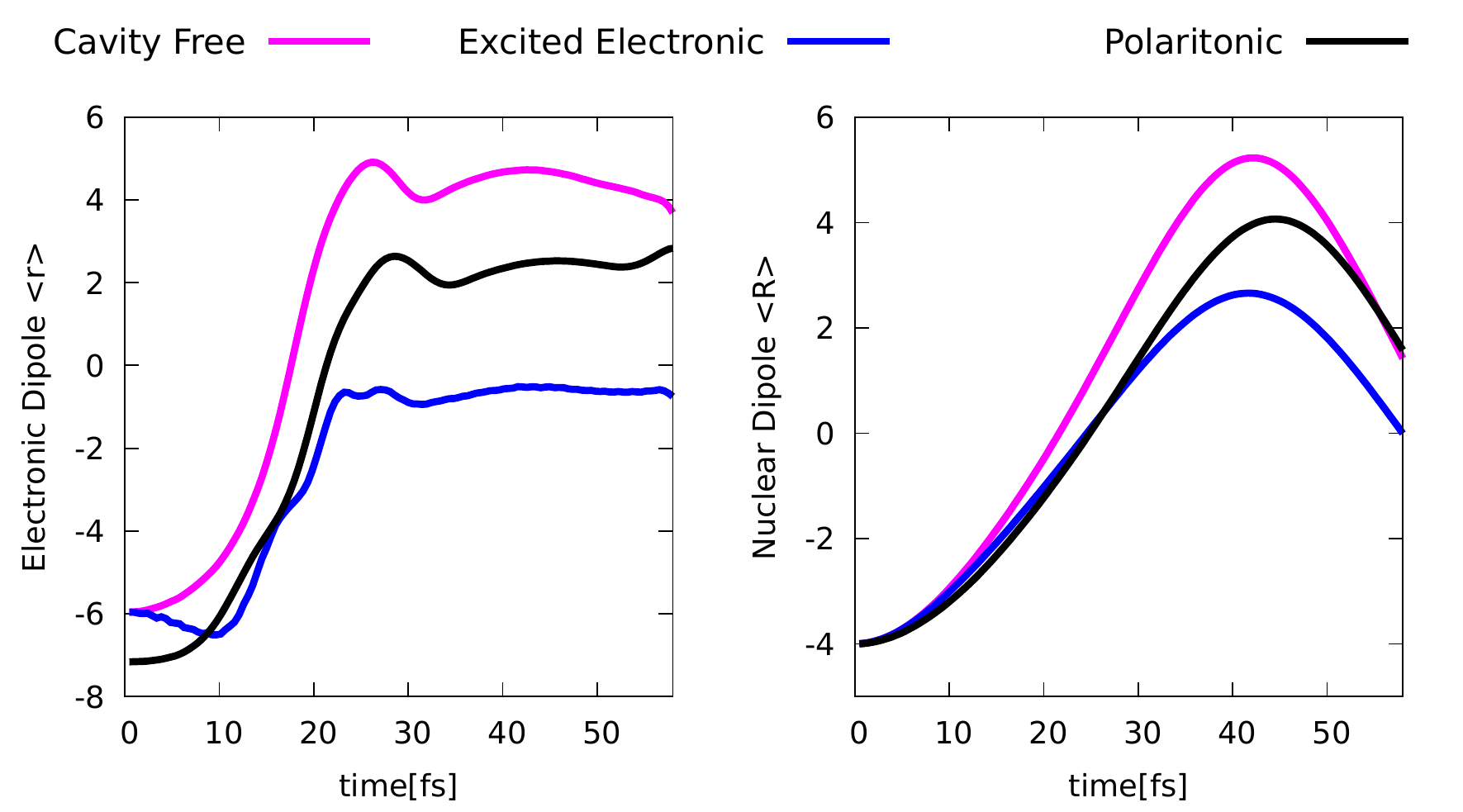}
\caption{Cavity-induced PCET suppression: Electronic and nuclear dipoles: cavity-free, in-cavity with the 0-photon excited electronic initial state, in-cavity with the polaritonic initial state.}
\label{fig:pcet_dipoles}
\end{figure}

Fig.~\ref{fig:pcet_dipoles} shows the electronic and nuclear dipoles (i.e. the expectation values of the electronic and nuclear coordinates respectively) when beginning in the first polaritonic state, compared with beginning in the 0-photon, first-excited electronic state which we will henceforth call the factorized state, and the cavity-free dipoles. We observe that PCET is suppressed also in the case of beginning in the polaritonic state, although not as much as in the factorized state.

\begin{figure}
\includegraphics[width=1.0\columnwidth]{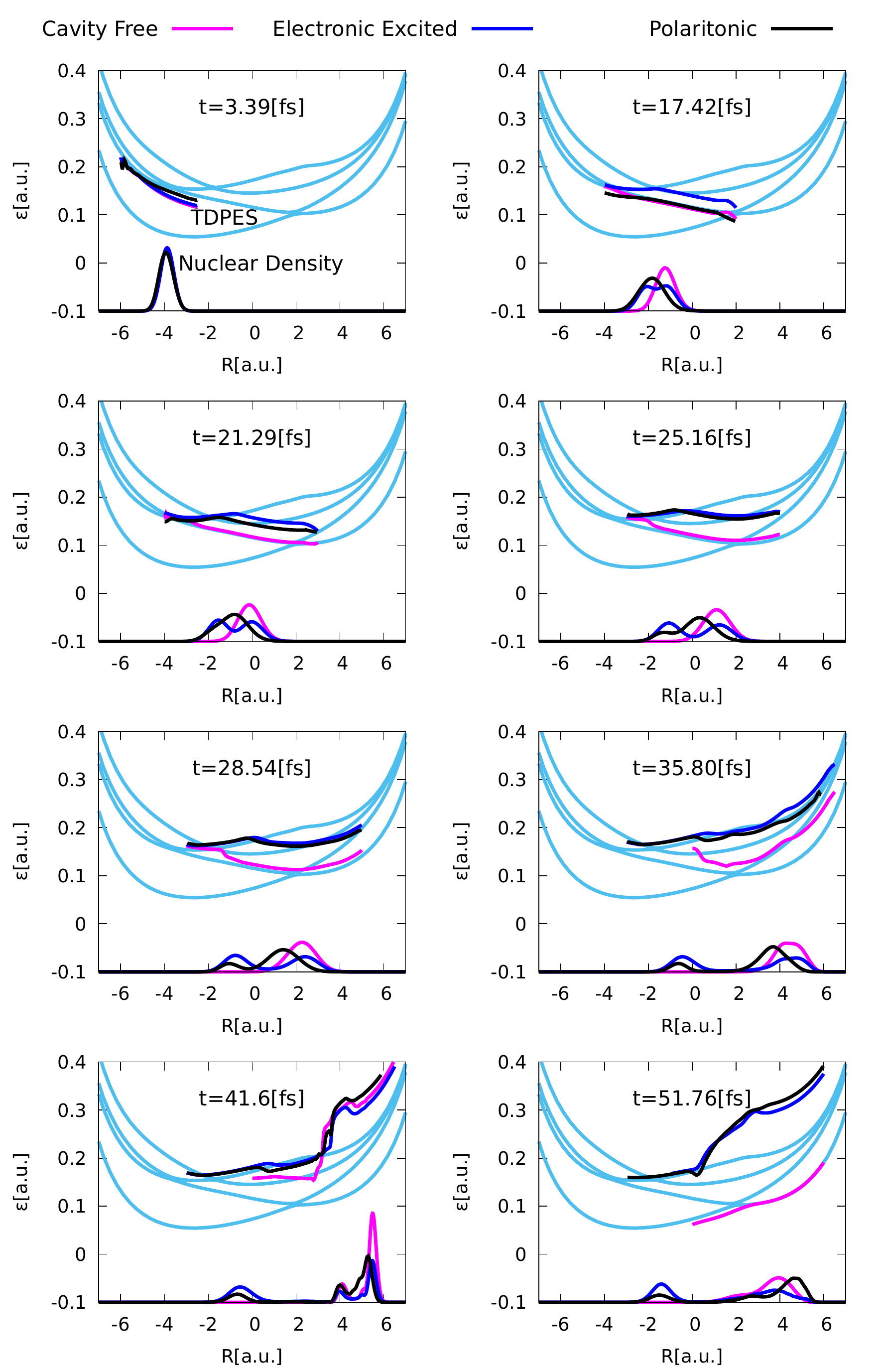}
\caption{Cavity-induced PCET suppression: Snapshots of the TDPES and nuclear density at the indicated times: cavity-free (pink), in-cavity with the 0-photon excited electronic initial state (dark blue), in-cavity with the polaritonic initial state (black). The backdrop of static polaritonic surfaces are shown in light blue for reference. }
\label{fig:pcet_tdpes_n}
\end{figure}

Snapshots of the TDPES are plotted in Fig.~\ref{fig:pcet_tdpes_n} along with the nuclear density, in comparison with the cavity-free case, and the factorized initial state case. 
Initially, the exact TDPES coincides with the polaritonic surface in the case of the polaritonic initial state, and along the BO excited surface in the case of the factorized state. 
Here we note that a global shift to the TDPES has no effect on the dynamics. As we do not set a zero of the potential at any particular $R$, when we say the TDPES coincides with a polaritonic surface, we really mean it is parallel to it. For $R$ near -4, the polaritonic surface has the character of the $1$-photon ground electronic BO state to the left and the character of the $0$-photon electronically excited BO state to the right. 
The gentler slope of the polaritonic surface on the left at early times slows the nuclear wavepacket relative to both the dynamics of the factorized state and the cavity-free dynamics. The soft kink in the exact TDPES in the
factorized initial-state case which separates a region where the surface slopes down to the left on the left and down to the right on the right ($R \approx -2$ at t = 17.42 fs), develops at somewhat later times in the case of the polaritonic initial state (visible in this case at $21.29$ fs). This results in the density splitting into two parts later than for the initially factorized case, and ultimately less of the wavepacket is trapped in the well on the left, compared to the factorized case. 
As in Ref.~\cite{LHM19}, we point out that this behavior cannot be predicted from the polaritonic surfaces alone; the polaritonic surfaces are independent of the choice of the initial state, and alone could not predict the distinct behavior resulting from different initial states. On the other hand, the distinct shapes of the TDPES for these two states does correlate with the different nuclear wavepacket dynamics, and their differences with the cavity-free TDPES also correlates with the different dynamics seen for the cavity-free nuclear wavepacket. Similar to the factorized case, the part of the nuclear wavepacket to the left of the soft kink is unable to overcome the energy barrier in the TDPES to reach the region of electron-nuclear avoided crossing ($\approx 2$a.u.) that leads to the electron transfer, thus resulting in suppression of PCET. Since less of the nuclear wavepacket is trapped on the left in the case of the initially polaritonic state because of the shape of the TDPES at the earlier times (gentler slope and later development of the soft kink), the suppression of PCET is less than in the factorized initial state case. 

 Underlying the shapes of the TDPES are the absorption and emission of a photon and electronic transitions between BO states which, through the conditional wavefunction dependence of the TDPES, control its structure. It can be instructive to consider these conditional quantities when interpreting the dynamics.  Movies of the TDPES and its components along with the density are given in the Supplemental information, as well as the $R$-resolved BO electronic populations $\vert C_i(R,t)\vert^2$ defined through 
\ben
\vert C_i(R,t)\vert^2 = \int\vert \langle \Phi_R^{{\rm BO},i} \vert \Psi\rangle_r\vert^2 dq
\label{eq:BOpops}
\een
 Movies of the $n$-photon resolved nuclear densities, $\chi^{n-\rm{ph}}(R,t) = \langle\xi_n\vert\Psi(t)\rangle_{r,q}$ where $\xi_n(q)$ are the harmonic oscillator eigenstates of the photonic Hamiltonian are also provided. 
 For the factorized  initial state at early times partial photon emission occurs throughout the wavepacket, while for the polaritonic initial state only the right part of the wavepacket becomes correlated with photon emission and electronic de-excitation while the left part becomes correlated with photon absorption and electronic excitation.  As the wavepacket moves towards the right, leaving the resonant region,  there has been less time for the part of wavepacket originally on the left to re-emit the photon it absorbed, so more remains on the excited electronic surface compared to the factorized case, and less ends up being trapped.

\begin{figure}
\includegraphics[width=1.0\columnwidth]{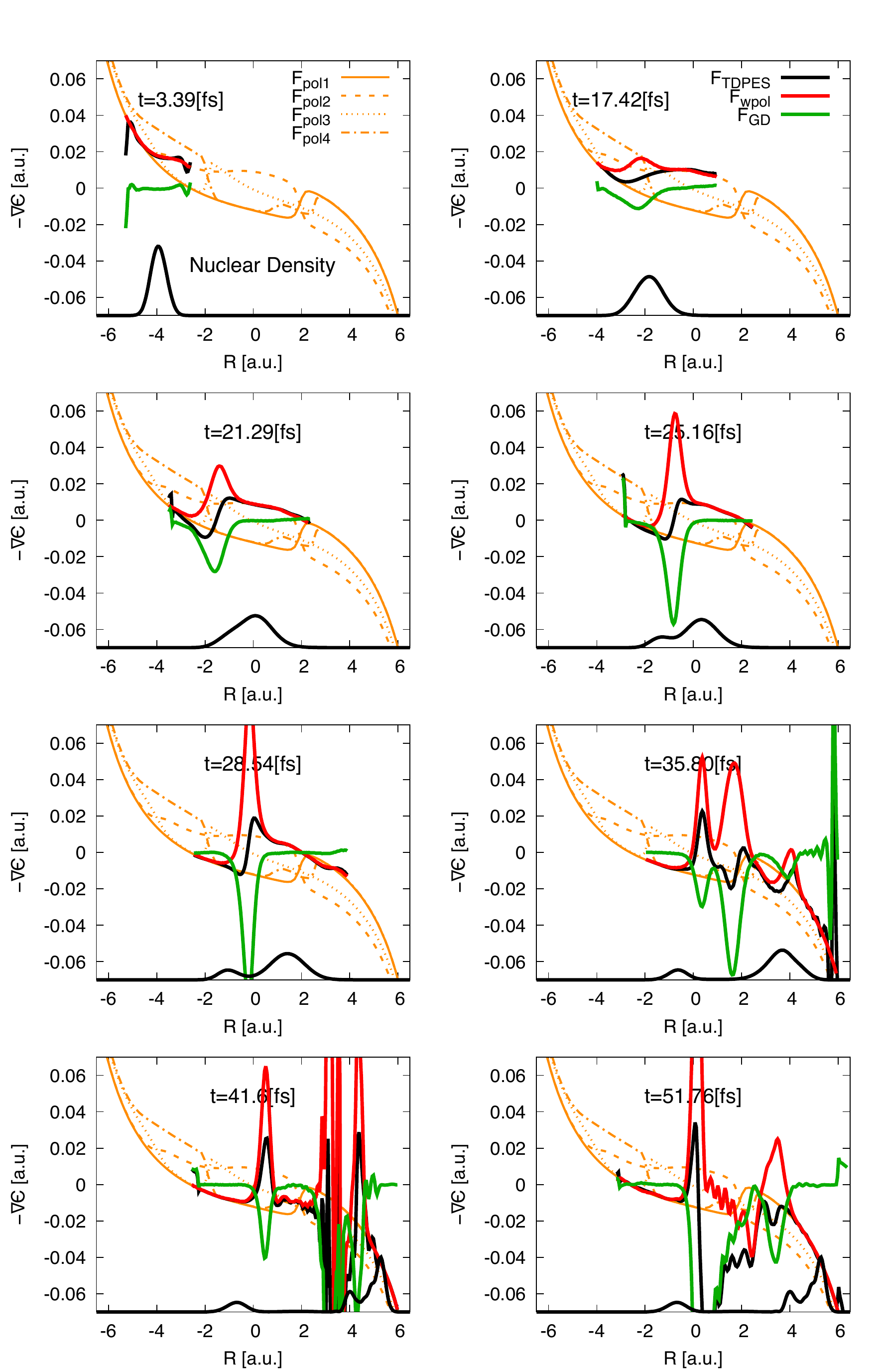}
\caption{Cavity-induced PCET suppression: Snapshots of the components of the force $-\nabla\eps$  (black) and its components coming from the weighted-polaritonic $-\nabla\eps\wpol$ (red) and the gauge-dependent $-\nabla\eps\GD$ (green)  components along with the density, when starting in the initial polaritonic state.   The gradients of the four lowest static polaritonic surfaces are shown for reference in orange. }
\label{fig:pcet_forces}
\end{figure}

We have found that the essential features of the surface are similar to what was found for the factorized case but details and timings are different. The gradients of the two dominant contributions to the TDPES, $\eps\wpol$ (Eq.~(\ref{eq:wpol})) and $\eps\GD$ (Eq.~(\ref{eq:GD})) are shown in Fig.~\ref{fig:pcet_forces}. At the first time shown, $3.39$fs, the force from the TDPES coincides with the slope of the second polaritonic surface as expected, with only the $\eps\wpol$ component contributing, but after very short times, both components begin to come into play. 
Where different parts of the nuclear wavepacket show a single polaritonic character that differs in different spatial regions, the gradient of $\eps\wpol$ follows the gradient of the relevant polaritonic surfaces in a piecewise manner,  with a peak in the region where they change character, e.g. at times $17.42, 21.29$, and $25.56$ fs. The gradient of $\eps\wpol$ is not simply a weighted average of the gradients of the polaritonic surfaces because the coefficients ${\cal{C}}_i(R,t)$ in Eq.~(\ref{eq:cond-exp}) vary with $R$, and in particular, these can vary significantly when the nuclear wavepacket correlated with one surface in one region switches to being correlated with another surface. The step-like transition in the coefficient leads to a peak-like structure in its gradient that affects the force the nuclei feel in this region. This structure tends to be partially compensated by a countering step in $\eps\GD$, whose gradient is zero in regions where the nuclear wavepacket is dominated by one surface; the total force from both terms becomes more a step-like feature in the transition region. Typically though, both components need to be considered through most of the nuclear wavepacket. For example, at $21.29$fs, the transition region is quite large, and the $\eps\GD$ correction to $\eps\wpol$ results in quite a different force throughout the nuclear density except for the right-most part, even reversing the sign of the force. As the nuclear wavepacket separates into two distinct parts evolving on different surfaces the opposing peaks in $\eps\GD$ and $\eps\wpol$, albeit large, have little effect when located in regions of low density in between the parts of the wavepacket (e.g. between $R \approx 0$ and $2$ at time $35.80$ fs). At this time we see the gradient of the total surface coincides with the 4th polaritonic surface (parallel to the lowest polaritonic surface, with character 1 photon and ground electronic state) in the extreme left part of the wavepacket, and is coming entirely from $\eps\wpol$ there. On the other hand the right-part of the wavepacket has a more mixed character, and really only on its extreme right is it dominated by a force from one polaritonic surface(the lowest surface). 
At later times, $\eps\GD$ plays a more complex role. For example, consider a situation where there is mixed character of the nuclear wavepacket such that  part is correlated with one polaritonic surface, while another part in the same region is correlated with a different polaritonic surface. These two surfaces could have quite different slopes, and $\eps\wpol$ typically resembles neither of them. This is particularly noticeable at the later times shown, where for $R > 0$ reflections from the right-hand-side of the polaritonic surfaces yield complex overlaps of the component densities, and $\eps\GD$ can dominate the force in some regions. It sometimes ensures a reflection which would not happen from purely $\eps\wpol$. 



To further demonstrate the importance of $\eps\GD$, and with a view to the development of mixed quantum-classical methods based on the TDPES, we run quasiclassical dynamics on the exact TDPES and compare with such dynamics on just the weighted-polaritonic component. That is, we first sample the initial Wigner phase-space distribution corresponding to the initial Gaussian wavepacket to find 3000 initial positions and momenta~\cite{H76}. (In fact the results were well-converged with just 2000 trajectories). We then propagate each of these using classical Hamilton's equations, $\dot{R} = P, \dot{P} = -\nabla\eps(R,t)/M$ and make a histogram of the resulting positions. 
Snapshots of the results are plotted in Fig.~\ref{fig:pcet_QC}, and a movie is provided in the Supplemental Information.   On the left are the histograms when the trajectories are classically propagated using the exact TDPES, and we observe the results are remarkably accurate. There appears to be two small differences: the ``trapped" part of the quasiclassical evolution appears to travel just a little further to the right, and at later times, the reflection is somewhat too enthusiastic. Still, given that such an evolution lacks adiabatic interference effects, quantum reflection and tunneling, and incurs errors when potentials are far from harmonic~\cite{H76}, the results here are very good. 
(We note that non-adiabatic interference has been shown to be captured by such an evolution~\cite{CAG16} but this is not a feature of the dynamics considered here). 
In contrast, on the right are the histograms when instead only the gauge-invariant $\eps\wpol$ is used for the force on the classical trajectories. The agreement is much worse, not surprising from the discussion of Fig.~\ref{fig:pcet_forces}; without the structures of $\eps\GD$ the wavepacket splitting is hardly captured, nor is the detailed reflection at later times. The wavepacket propagates more or less in one piece to the right until it reaches the sharp rise of all the surfaces when the trajectories slow down before turning around.

\begin{figure}
\includegraphics[width=\columnwidth]{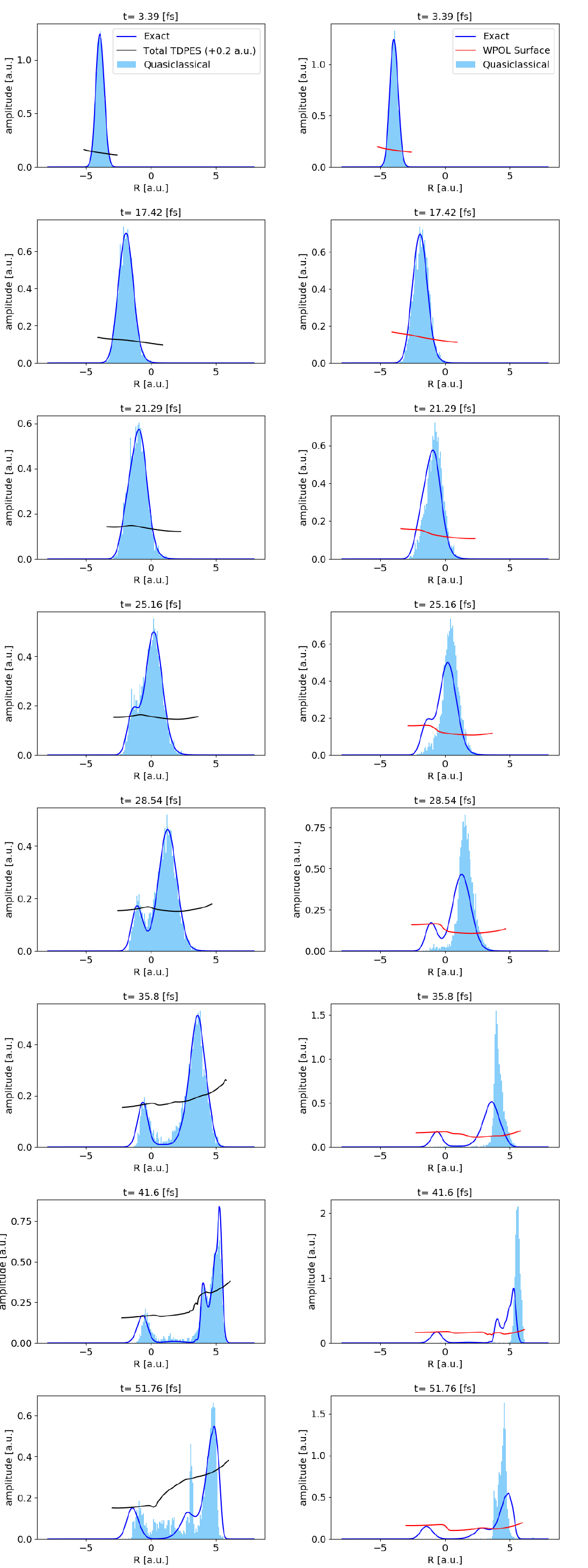}
\caption{Cavity-induced PCET suppression: Snapshots of the density resulting from quasiclassical propagation on the exact TDPES (left) and on the weighted polaritonic surface (right), when starting in the initial polaritonic state.}
\label{fig:pcet_QC}
\end{figure}

\subsection{Cavity-induced  ELEX}
\label{sec:elex}
We now turn to the second set of parameters for our model. 

Our initial state is again Gaussian in the nuclear variables, centered at $R = -4$a.u., and placed on the upper electronic surface, with zero photons in the cavity: $\Psi(r,R,q,t) = {\cal{N}}e^{-2.85(R+4)^2}\Phi_R^{\rm BO}(r)\xi^{(0)}(q)$ where $\xi^{(0)} = (\omega_c/\pi)^{1/4}e^{-\omega_c q^2/2}$  is the zero-photon state in the cavity.  We choose our cavity-resonance at $\omega_c = 0.049$a.u., resonant with the BO frequency between the ground and first excited electronic states at $R = \pm2.2$a.u., and also resonant between the ground and 2nd excited BO states at $R = 0$a.u (see right-hand panel of Fig.~\ref{fig:bopol}). 

\begin{figure}
\includegraphics[width=1.0\columnwidth]{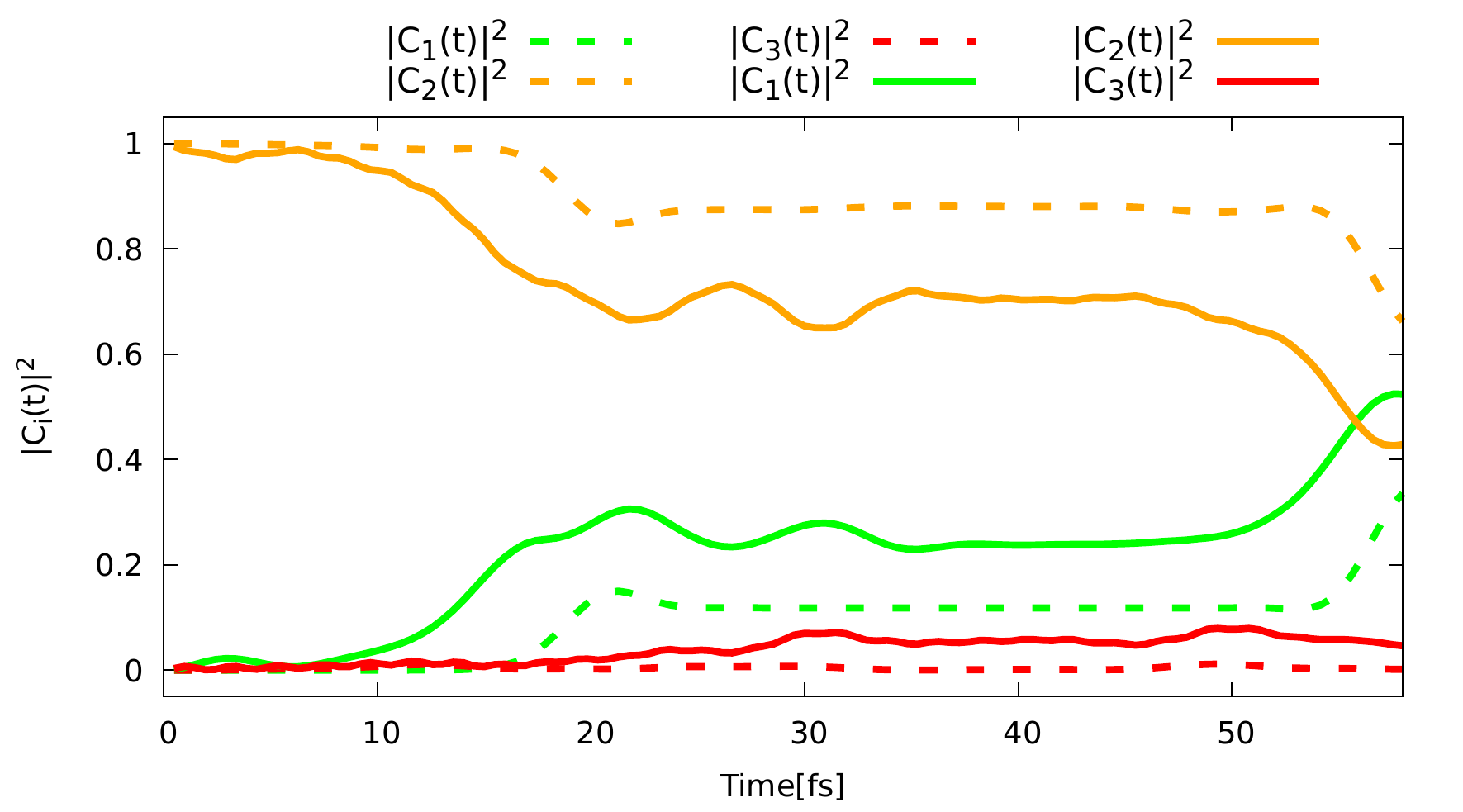}
\caption{Cavity-induced ELEX: BO populations, $\vert C_i(t)\vert^2$ as a function of time, for the lowest three states, in the in-cavity (solid) and cavity-free (dashed) cases.}
\label{fig:elex}
\end{figure}

The dynamics outside the cavity is relatively simple: the nuclear wavepacket slides down the first excited BO surface, meeting a weak avoided crossing at the origin, where a little population is transferred to the ground-state. A second transfer occurs later upon reflection back to this region.  Figure~\ref{fig:elex} shows the BO populations, $\vert C_i(t)\vert^2$, where  $\vert C_i(t)\vert^2 = \int dR \vert C_i(R,t)\vert^2 $ with $C_i(R,t)$ defined in Eq.~(\ref{eq:BOpops}).
There is a very small transfer from the first to the second excited state, only just discernible in the figure. 
Now when the molecule is placed in the cavity, we see a more rapid and larger transfer to the ground-state surfaces along with a much greater rise in the population of the second excited BO state. 
Thus, this is an example of cavity-induced electronic excitation. 

The phenomena can be only partially understood from the polaritonic surfaces shown in Fig.~\ref{fig:bopol} because these surfaces look practically identical to the eye for much smaller coupling strengths where the dynamics is almost the same as outside the cavity. Still, the fact that both the ground-state population and the 2nd excited population are enhanced at some times in the wavepacket evolution  can be anticipated from the avoided crossings in the polaritonic surfaces and the nature of these surfaces: the  wavepacket starts on the 2nd Born-Oppenheimer surface coinciding with the 3rd polaritonic surface at $R = -4$a.u., approaches the narrow avoided crossing with the 2nd polaritonic surface around $R = -2.2$a.u. suggesting a partial transfer to this surface. The part remaining on the 3rd surface is associated with a 1-photon ground-electronic character to the right of the crossing, as is evident from the shape of that surface, aligning with the ground-state surface shifted up by the energy of one photon in that region. This transfer is consistent with the earlier transfer of population from the electronically excited to the ground electronic state compared with the cavity-free case. The enhancement of the 3rd electronic state population then arises when the part of the wavepacket on the third polaritonic surface (that has the 1-photon ground electronic state character) approaches the narrow avoided crossing with the 4th polaritonic state at $R = 0$. Moving away from the crossing this 4th surface has the character of the 3rd electronic state. 

\begin{figure}
\includegraphics[width=1.0\columnwidth]{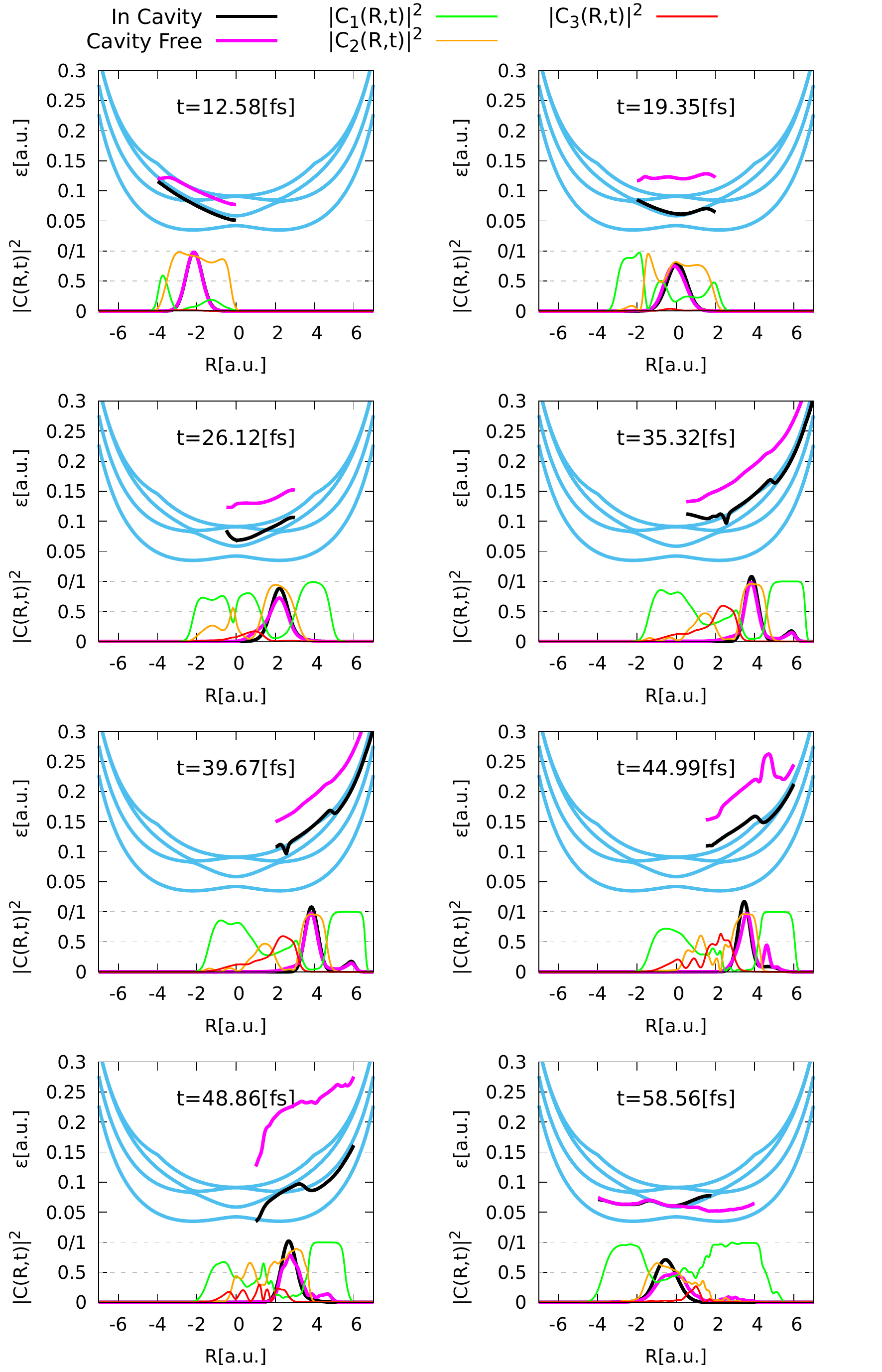}
\caption{Cavity-induced ELEX: Snapshots of the TDPES and nuclear density for the case of the induced electronic excitation: cavity-free (pink), in-cavity with the factorized initial state (black), against the background of polaritonic surfaces (pale blue)}
\label{fig:elex_tdpes_n}
\end{figure}

We now turn to the exact TDPES for this problem, shown at time-snapshots in Fig.~\ref{fig:elex_tdpes_n} along with the cavity-free surface. The effect of the cavity on the nuclear wavepacket dynamics is much smaller than in the PCET example except at later times, but to highlight the electronic character, we plot here also the $R$-resolved electronic populations, defined in Eq.~(\ref{eq:BOpops}). These are plotted as the thin lines on the split scale shown at the bottom of each panel. We see the growth of the coefficient of the 3rd electronic state (red) beginning  as the wavepacket goes through $R = 0$ as expected, and this coefficient is carried along in the tail of the wavepacket as it evolves to the right and then reflects. We see a very complex character develop by the end, especially in the left hand tail of the wavepacket which has a complex and oscillatory  electronic structure. Again, the TDPES shows a piecewise polaritonic character when there are distinct single-surface dominated regions, but  again in regions of mixed character, the polaritonic surface structure does not help to predict the correct force as given by the TDPES. 

\begin{figure}
\includegraphics[width=1.0\columnwidth]{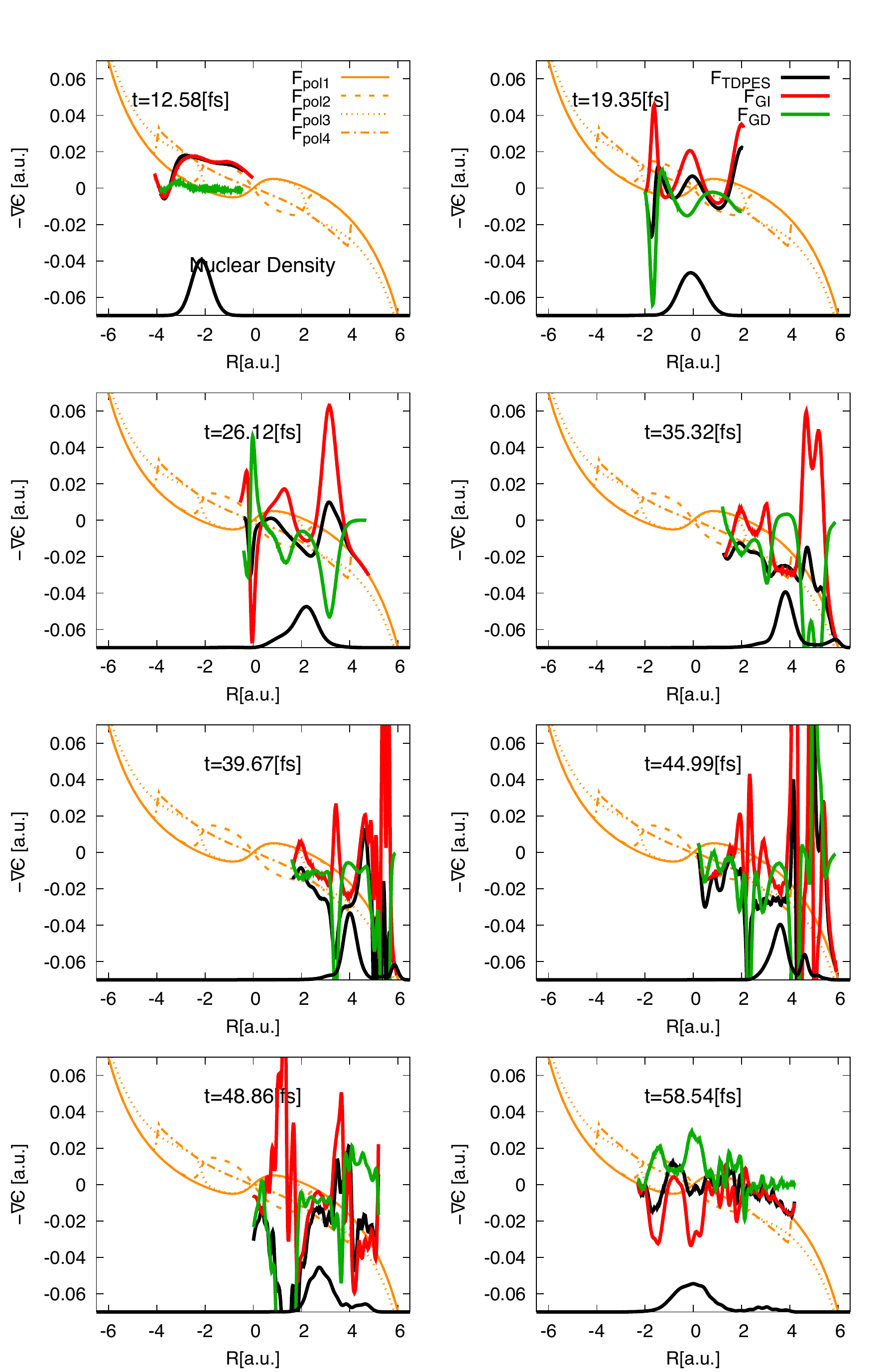}
\caption{Cavity-induced ELEX: Snapshots of the components of the force $-\nabla\eps$ and its components coming from the weighted-polaritonic ($-\nabla\eps\wpol$) and the gauge-dependent  ($-\nabla\eps\GD$) components, along with the density. The orange lines in the backdrop are the gradients of the four lowest polaritonic surfaces. }
\label{fig:elex_forces}
\end{figure}

\begin{figure}
\includegraphics[width=\columnwidth]{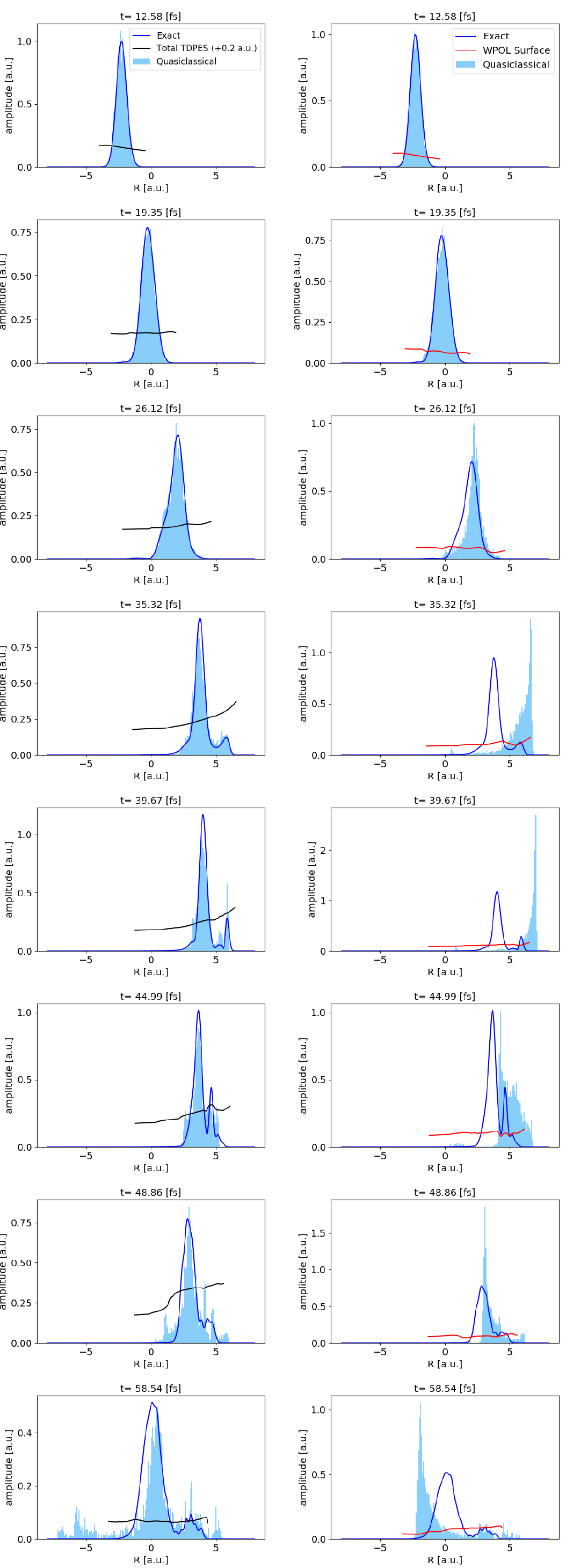}
\caption{Cavity-induced ELEX suppression: Snapshots of the density resulting from quasiclassical propagation on the exact TDPES (left) and on the weighted polaritonic surface (right).  }
\label{fig:elex_QC}
\end{figure}

Similar observations as for the PCET case can be made for the force plots in Fig.~\ref{fig:elex_forces}, with the salient point being that the force from the $\eps\wpol$ is, more often than not, countered by an opposing force from $\eps\GD$ in the mixed regions, and does not resemble a weighted average of forces from polaritonic surfaces.  Once again the quasiclassical propagation on the exact TDPES in Figure~\ref{fig:elex_QC} performs remarkable well, while the importance of including the force from $\eps\GD$ is evident in its poorer prediction of the dynamics in the right-hand panel of the figure.

\section{Conclusions and Outlook}
The exact potentials of the EF studied here provide the complete information to propagate the nuclei in cavity-QED problems.  Our examples show how their structure correlates directly with the dynamics of the nuclear wavepacket, for two different cavity-altered phenomena. Unlike polaritonic surfaces, these exact potentials distinguish initial states of different polaritonic character, and reflect, for example, the different degrees of suppression achieved when an initially polaritonic state is used compared with an initially pure electronic excitation.
We note that other possible factorizations may lead to interesting insights, for example, taking the marginal factor to include both the photonic and nuclear degrees of freedom, with the conditional electronic wavefunction parametrically dependent on both degrees of freedom, $\chi(R,q,t)\Phi_{R,q}(r,t)$. The resulting surfaces would provide corrections to the sometimes used ``cavity-BO" surfaces \cite{FARR17}.
For real molecules of interest, finding the exact potentials of the EF method is as hard as solving the full TDSE exactly,  so instead approximations need to be made. By studying the exact features in models where they can be found, such as the ones presented here, insight is gained into the effect that the various terms have on the nuclear dynamics  and the results provide guidance for building approximate propagation methods. Similar features have been found in the exact surfaces driving the nuclei when coupled to electrons in the absence of cavities or external fields~\cite{AASG13,AASMMG15} which have then motivated and justified the mixed quantum-classical propagation methods of Ref.~\cite{MAG15,AMAG16,MATG17,CAT18} and Ref~\cite{HLM18,FMK19,FPMC19,FPMK18}. This suggests that mixed quantum-classical methods along these lines can be generalized to include photons. The polaritonic surfaces play the role that the underlying Born-Oppenheimer surfaces would in that situation. 

In such mixed quantum-classical methods, a key issue is to determine the force on the classical nuclei. While the Ehrenfest method takes a weighted-average over the Born-Oppenheimer surfaces, trajectory surface-hopping uses the gradient of a single surface at any time, switching between surfaces according to a stochastic algorithm. If applied now to problems in polaritonic chemistry, our results here stress that within an Ehrenfest-like approach, a weighted-average of polaritonic surfaces would lead to significantly incorrect dynamics, while if the gauge-dependent term was included to form the total TDPES, the results would improve significantly. However it is important to realize that the results shown here for the quasiclassical propagation on the weighted polaritonic surfaces are at a higher level than a traditional Ehrenfest calculation, since the latter involves an expansion of the conditional electronic wavefunction with nuclear-coordinate-independent coefficients, while the weights used in our calculations include coordinate dependence. The coordinate-dependence of the weights affects the forces obtained from the gradients as we explicitly showed. That is, the forces in a traditional Ehrenfest calculation differ from those from the exact weighted-polaritonic surface even if the populations of the polaritonic  states averaged over the system are the same. 

A traditional surface-hopping scheme, on the other hand, would be even more dogged by the problems of coherence than in the usual cavity-free case, due to the increased number of avoided amd trivial crossings between the polaritonic surfaces compared to the Born-Oppenheimer ones \cite{GCFMT2019}. Both decoherence and recoherence effects would need to be captured for an accurate determination of the dynamics. The EF-based surface-hopping approach of Ref.~\cite{HLM18,FMK19,FPMC19,FPMK18} can be generalized to the polaritonic case. Here, an additional term, derived from the EF, appears in the electronic equation of motion and how well it captures these effects will be explored in future work, as will other methods of developing EF-based mixed quantum-classical methods for polaritonic chemistry. 

\section*{Supplementary Material}
Movies of the dynamics are provided in the supplementary material as follows: 
\begin{enumerate}
    \item \textbf{ Movie Fig3.mp4} Upper row from left to right:
$\epsilon(R, t)$, $\epsilon\wpol(R, t)$, $\epsilon\GD(R, t)$ for cavity free (pink), in-cavity with the 0-photon excited electronic initial state (dark blue), in-cavity with the polaritonic initial state (black). The backdrop of static polaritonic surfaces are shown in light blue for reference (as in Fig. \ref{fig:pcet_tdpes_n}). Lower row from left to right: 0-photon resolved density and 1-photon resolved density in the same color code as upper row, and the $R$-resolved Born-Oppenheimer population for (a) in-cavity with the polaritonic initial state (b) in-cavity with the 0-photon excited electronic initial state (c) cavity-free with $|C_1(R, t)|^2$ (green), $|C_2(R, t)|^2$ (orange) and $|C_3(R, t)|^2$ (red) (same color code as Fig. \ref{fig:elex_tdpes_n}). 

\item \textbf{ Movie Fig5.mp4}
Left Panel: Quasiclassical propagation of 3000 trajectories (orange) on the total TDPES $\epsilon(R,t)$ for the cavity- induced suppression of PCET (case IIIA). These are plotted as a function of $R(a.u.)$ against exact solution (blue), Born- Oppenheimer ground state population (red) and Born-Oppenheimer first-excited state population (green). Right Panel: Same as above, propagated using the weighted-polaritonic PES $\epsilon\wpol(R, t)$.

\item \textbf{ Movie Fig7.mp4} 
Upper row from left to right: $\epsilon(R, t)$, $\epsilon\wpol(R, t)$, $\epsilon\GD(R, t)$ for cavity free (pink), in-cavity with the 0-photon excited electronic initial state (black). The backdrop of static polaritonic surfaces are shown in light blue for reference (as in Fig. \ref{fig:elex_tdpes_n}) Lower row from left to right: 0-photon resolved density and 1-photon resolved density in the same color code as upper row, and the $R$-resolved Born-Oppenheimer Population for (a) with the 0-photon excited electronic initial state and (b) cavity-free with $|C_1(R, t)|^2$ (green), $|C_2(R, t)|^2$ (orange) and $|C_3(R, t)|^2$ (red) (same color code as figure \ref{fig:elex_tdpes_n}).

\item \textbf{ Movie Fig9.mp4}
Left Panel: Quasiclassical propagation of 3000 trajectories (orange) on the total TDPES $\epsilon(R,t)$ for the cavity- induced electronic excitation (case IIIB). Plotted as a function of $R(a.u.)$ against exact solution (blue), Born- Oppenheimer ground state population (red) and Born-Oppenheimer first-excited state population (green). Right Panel: Same as above, propagated using the weighted-polaritonic PES $\epsilon\wpol(R, t)$.

\end{enumerate}

\begin{acknowledgments}{Financial support from the US National Science Foundation
CHE-1940333 (N.T.M, N.M.H, and B.R. ), the Department of Energy, Office
of Basic Energy Sciences, Division of Chemical Sciences,
Geosciences and Biosciences under Award DE-SC0020044(L.L.), and the RISE program at Hunter College, Grant No. 5R25GM060665-20
are gratefully acknowledged. }
\end{acknowledgments}

\section{Data Availability Statement} 
The data that support the findings of this study are available from the corresponding author upon reasonable request.
\bibliography{./ref_na}

\begin{thebibliography}{57}%
\makeatletter
\providecommand \@ifxundefined [1]{%
 \@ifx{#1\undefined}
}%
\providecommand \@ifnum [1]{%
 \ifnum #1\expandafter \@firstoftwo
 \else \expandafter \@secondoftwo
 \fi
}%
\providecommand \@ifx [1]{%
 \ifx #1\expandafter \@firstoftwo
 \else \expandafter \@secondoftwo
 \fi
}%
\providecommand \natexlab [1]{#1}%
\providecommand \enquote  [1]{``#1''}%
\providecommand \bibnamefont  [1]{#1}%
\providecommand \bibfnamefont [1]{#1}%
\providecommand \citenamefont [1]{#1}%
\providecommand \href@noop [0]{\@secondoftwo}%
\providecommand \href [0]{\begingroup \@sanitize@url \@href}%
\providecommand \@href[1]{\@@startlink{#1}\@@href}%
\providecommand \@@href[1]{\endgroup#1\@@endlink}%
\providecommand \@sanitize@url [0]{\catcode `\\12\catcode `\$12\catcode
  `\&12\catcode `\#12\catcode `\^12\catcode `\_12\catcode `\%12\relax}%
\providecommand \@@startlink[1]{}%
\providecommand \@@endlink[0]{}%
\providecommand \url  [0]{\begingroup\@sanitize@url \@url }%
\providecommand \@url [1]{\endgroup\@href {#1}{\urlprefix }}%
\providecommand \urlprefix  [0]{URL }%
\providecommand \Eprint [0]{\href }%
\providecommand \doibase [0]{http://dx.doi.org/}%
\providecommand \selectlanguage [0]{\@gobble}%
\providecommand \bibinfo  [0]{\@secondoftwo}%
\providecommand \bibfield  [0]{\@secondoftwo}%
\providecommand \translation [1]{[#1]}%
\providecommand \BibitemOpen [0]{}%
\providecommand \bibitemStop [0]{}%
\providecommand \bibitemNoStop [0]{.\EOS\space}%
\providecommand \EOS [0]{\spacefactor3000\relax}%
\providecommand \BibitemShut  [1]{\csname bibitem#1\endcsname}%
\let\auto@bib@innerbib\@empty
\bibitem [{\citenamefont {Purcell}(1946)}]{Purcell1946}%
  \BibitemOpen
  \bibfield  {author} {\bibinfo {author} {\bibfnamefont {E.~M.}\ \bibnamefont
  {Purcell}},\ }\href@noop {} {\bibfield  {journal} {\bibinfo  {journal} {Phys.
  Rev.}\ }\textbf {\bibinfo {volume} {69}},\ \bibinfo {pages} {681} (\bibinfo
  {year} {1946})}\BibitemShut {NoStop}%
\bibitem [{\citenamefont {Ribeiro}\ \emph {et~al.}(2018)\citenamefont
  {Ribeiro}, \citenamefont {Martinez-Martinez}, \citenamefont {Du},
  \citenamefont {Campos-Gonzalez-Angulo},\ and\ \citenamefont
  {Yuen-Zhou}}]{RMDCY18}%
  \BibitemOpen
  \bibfield  {author} {\bibinfo {author} {\bibfnamefont {R.~F.}\ \bibnamefont
  {Ribeiro}}, \bibinfo {author} {\bibfnamefont {L.~A.}\ \bibnamefont
  {Martinez-Martinez}}, \bibinfo {author} {\bibfnamefont {M.}~\bibnamefont
  {Du}}, \bibinfo {author} {\bibfnamefont {J.}~\bibnamefont
  {Campos-Gonzalez-Angulo}}, \ and\ \bibinfo {author} {\bibfnamefont
  {J.}~\bibnamefont {Yuen-Zhou}},\ }\href@noop {} {\bibfield  {journal}
  {\bibinfo  {journal} {Chem. Sci.}\ }\textbf {\bibinfo {volume} {9}},\
  \bibinfo {pages} {6325} (\bibinfo {year} {2018})}\BibitemShut {NoStop}%
\bibitem [{\citenamefont {Herrera}\ and\ \citenamefont
  {Owrutsky}(2020)}]{HO20}%
  \BibitemOpen
  \bibfield  {author} {\bibinfo {author} {\bibfnamefont {F.}~\bibnamefont
  {Herrera}}\ and\ \bibinfo {author} {\bibfnamefont {J.}~\bibnamefont
  {Owrutsky}},\ }\href {\doibase 10.1063/1.5136320} {\bibfield  {journal}
  {\bibinfo  {journal} {The Journal of Chemical Physics}\ }\textbf {\bibinfo
  {volume} {152}},\ \bibinfo {pages} {100902} (\bibinfo {year}
  {2020})}\BibitemShut {NoStop}%
\bibitem [{\citenamefont {Feist}\ \emph {et~al.}(2018)\citenamefont {Feist},
  \citenamefont {Galego},\ and\ \citenamefont {Garcia-Vidal}}]{FGG18}%
  \BibitemOpen
  \bibfield  {author} {\bibinfo {author} {\bibfnamefont {J.}~\bibnamefont
  {Feist}}, \bibinfo {author} {\bibfnamefont {J.}~\bibnamefont {Galego}}, \
  and\ \bibinfo {author} {\bibfnamefont {F.~J.}\ \bibnamefont {Garcia-Vidal}},\
  }\href@noop {} {\bibfield  {journal} {\bibinfo  {journal} {ACS Photonics}\
  }\textbf {\bibinfo {volume} {5}},\ \bibinfo {pages} {205} (\bibinfo {year}
  {2018})}\BibitemShut {NoStop}%
\bibitem [{\citenamefont {Ruggenthaler}\ \emph {et~al.}(2018)\citenamefont
  {Ruggenthaler}, \citenamefont {Tancogne-Dejean}, \citenamefont {Flick},
  \citenamefont {Appel},\ and\ \citenamefont {Rubio}}]{RTFAR18}%
  \BibitemOpen
  \bibfield  {author} {\bibinfo {author} {\bibfnamefont {M.}~\bibnamefont
  {Ruggenthaler}}, \bibinfo {author} {\bibfnamefont {N.}~\bibnamefont
  {Tancogne-Dejean}}, \bibinfo {author} {\bibfnamefont {J.}~\bibnamefont
  {Flick}}, \bibinfo {author} {\bibfnamefont {H.}~\bibnamefont {Appel}}, \ and\
  \bibinfo {author} {\bibfnamefont {A.}~\bibnamefont {Rubio}},\ }\href@noop {}
  {\bibfield  {journal} {\bibinfo  {journal} {Nature Reviews Chemistry}\
  }\textbf {\bibinfo {volume} {2}},\ \bibinfo {pages} {0118} (\bibinfo {year}
  {2018})}\BibitemShut {NoStop}%
\bibitem [{\citenamefont {Hertzog}\ \emph {et~al.}(2019)\citenamefont
  {Hertzog}, \citenamefont {Wang}, \citenamefont {Mony},\ and\ \citenamefont
  {Borjesson}}]{HWMB19}%
  \BibitemOpen
  \bibfield  {author} {\bibinfo {author} {\bibfnamefont {M.}~\bibnamefont
  {Hertzog}}, \bibinfo {author} {\bibfnamefont {M.}~\bibnamefont {Wang}},
  \bibinfo {author} {\bibfnamefont {J.}~\bibnamefont {Mony}}, \ and\ \bibinfo
  {author} {\bibfnamefont {K.}~\bibnamefont {Borjesson}},\ }\href {\doibase
  10.1039/C8CS00193F} {\bibfield  {journal} {\bibinfo  {journal} {Chem. Soc.
  Rev.}\ }\textbf {\bibinfo {volume} {48}},\ \bibinfo {pages} {937} (\bibinfo
  {year} {2019})}\BibitemShut {NoStop}%
\bibitem [{\citenamefont {Kowalewski}\ \emph {et~al.}(2016)\citenamefont
  {Kowalewski}, \citenamefont {Bennett},\ and\ \citenamefont
  {Mukamel}}]{KBM16}%
  \BibitemOpen
  \bibfield  {author} {\bibinfo {author} {\bibfnamefont {M.}~\bibnamefont
  {Kowalewski}}, \bibinfo {author} {\bibfnamefont {K.}~\bibnamefont {Bennett}},
  \ and\ \bibinfo {author} {\bibfnamefont {S.}~\bibnamefont {Mukamel}},\
  }\href@noop {} {\bibfield  {journal} {\bibinfo  {journal} {The Journal of
  Physical Chemistry Letters}\ }\textbf {\bibinfo {volume} {7}},\ \bibinfo
  {pages} {2050} (\bibinfo {year} {2016})}\BibitemShut {NoStop}%
\bibitem [{\citenamefont {Szidarovszky}\ \emph {et~al.}(2018)\citenamefont
  {Szidarovszky}, \citenamefont {Halasz}, \citenamefont {Csaszar},
  \citenamefont {Cederbaum},\ and\ \citenamefont {Vibok}}]{SHCCV18}%
  \BibitemOpen
  \bibfield  {author} {\bibinfo {author} {\bibfnamefont {T.}~\bibnamefont
  {Szidarovszky}}, \bibinfo {author} {\bibfnamefont {G.~J.}\ \bibnamefont
  {Halasz}}, \bibinfo {author} {\bibfnamefont {A.~G.}\ \bibnamefont {Csaszar}},
  \bibinfo {author} {\bibfnamefont {L.~S.}\ \bibnamefont {Cederbaum}}, \ and\
  \bibinfo {author} {\bibfnamefont {A.}~\bibnamefont {Vibok}},\ }\href
  {\doibase 10.1021/acs.jpclett.8b01102} {\bibfield  {journal} {\bibinfo
  {journal} {The Journal of Physical Chemistry Letters}\ }\textbf {\bibinfo
  {volume} {9}},\ \bibinfo {pages} {2739} (\bibinfo {year} {2018})}\BibitemShut
  {NoStop}%
\bibitem [{\citenamefont {Baranov}\ \emph {et~al.}(2018)\citenamefont
  {Baranov}, \citenamefont {Wers\"all}, \citenamefont {Cuadra}, \citenamefont
  {Antosiewicz},\ and\ \citenamefont {Shegai}}]{BWCAS18}%
  \BibitemOpen
  \bibfield  {author} {\bibinfo {author} {\bibfnamefont {D.~G.}\ \bibnamefont
  {Baranov}}, \bibinfo {author} {\bibfnamefont {M.}~\bibnamefont {Wers\"all}},
  \bibinfo {author} {\bibfnamefont {J.}~\bibnamefont {Cuadra}}, \bibinfo
  {author} {\bibfnamefont {T.~J.}\ \bibnamefont {Antosiewicz}}, \ and\ \bibinfo
  {author} {\bibfnamefont {T.}~\bibnamefont {Shegai}},\ }\href@noop {}
  {\bibfield  {journal} {\bibinfo  {journal} {ACS Photonics}\ }\textbf
  {\bibinfo {volume} {5}},\ \bibinfo {pages} {24} (\bibinfo {year}
  {2018})}\BibitemShut {NoStop}%
\bibitem [{\citenamefont {Xiang}\ \emph {et~al.}(2020)\citenamefont {Xiang},
  \citenamefont {Ribeiro}, \citenamefont {Du}, \citenamefont {Chen},
  \citenamefont {Yang}, \citenamefont {Wang}, \citenamefont {Yuen-Zhou},\ and\
  \citenamefont {Xiong}}]{XRDCYWYZ20}%
  \BibitemOpen
  \bibfield  {author} {\bibinfo {author} {\bibfnamefont {B.}~\bibnamefont
  {Xiang}}, \bibinfo {author} {\bibfnamefont {R.~F.}\ \bibnamefont {Ribeiro}},
  \bibinfo {author} {\bibfnamefont {M.}~\bibnamefont {Du}}, \bibinfo {author}
  {\bibfnamefont {L.}~\bibnamefont {Chen}}, \bibinfo {author} {\bibfnamefont
  {Z.}~\bibnamefont {Yang}}, \bibinfo {author} {\bibfnamefont {J.}~\bibnamefont
  {Wang}}, \bibinfo {author} {\bibfnamefont {J.}~\bibnamefont {Yuen-Zhou}}, \
  and\ \bibinfo {author} {\bibfnamefont {W.}~\bibnamefont {Xiong}},\ }\href
  {\doibase 10.1126/science.aba3544} {\bibfield  {journal} {\bibinfo  {journal}
  {Science}\ }\textbf {\bibinfo {volume} {368}},\ \bibinfo {pages} {665}
  (\bibinfo {year} {2020})}\BibitemShut {NoStop}%
\bibitem [{\citenamefont {Hunter}(1974)}]{Hunter74}%
  \BibitemOpen
  \bibfield  {author} {\bibinfo {author} {\bibfnamefont {G.}~\bibnamefont
  {Hunter}},\ }\href@noop {} {\bibfield  {journal} {\bibinfo  {journal} {Int.
  J. Quantum Chem.}\ }\textbf {\bibinfo {volume} {8}},\ \bibinfo {pages} {413}
  (\bibinfo {year} {1974})}\BibitemShut {NoStop}%
\bibitem [{\citenamefont {Hunter}(1975)}]{Hunter75}%
  \BibitemOpen
  \bibfield  {author} {\bibinfo {author} {\bibfnamefont {G.}~\bibnamefont
  {Hunter}},\ }\href@noop {} {\bibfield  {journal} {\bibinfo  {journal} {Int.
  J. Quantum Chem.}\ }\textbf {\bibinfo {volume} {9}},\ \bibinfo {pages} {237}
  (\bibinfo {year} {1975})}\BibitemShut {NoStop}%
\bibitem [{\citenamefont {Abedi}\ \emph {et~al.}(2010)\citenamefont {Abedi},
  \citenamefont {Maitra},\ and\ \citenamefont {Gross}}]{AMG10}%
  \BibitemOpen
  \bibfield  {author} {\bibinfo {author} {\bibfnamefont {A.}~\bibnamefont
  {Abedi}}, \bibinfo {author} {\bibfnamefont {N.~T.}\ \bibnamefont {Maitra}}, \
  and\ \bibinfo {author} {\bibfnamefont {E.~K.~U.}\ \bibnamefont {Gross}},\
  }\href@noop {} {\bibfield  {journal} {\bibinfo  {journal} {Phys. Rev. Lett.}\
  }\textbf {\bibinfo {volume} {105}},\ \bibinfo {pages} {123002} (\bibinfo
  {year} {2010})}\BibitemShut {NoStop}%
\bibitem [{\citenamefont {Abedi}\ \emph {et~al.}(2012)\citenamefont {Abedi},
  \citenamefont {Maitra},\ and\ \citenamefont {Gross}}]{AMG12}%
  \BibitemOpen
  \bibfield  {author} {\bibinfo {author} {\bibfnamefont {A.}~\bibnamefont
  {Abedi}}, \bibinfo {author} {\bibfnamefont {N.~T.}\ \bibnamefont {Maitra}}, \
  and\ \bibinfo {author} {\bibfnamefont {E.~K.~U.}\ \bibnamefont {Gross}},\
  }\href@noop {} {\bibfield  {journal} {\bibinfo  {journal} {J. Chem. Phys.}\
  }\textbf {\bibinfo {volume} {137}},\ \bibinfo {pages} {22A530} (\bibinfo
  {year} {2012})}\BibitemShut {NoStop}%
\bibitem [{\citenamefont {Hoffmann}\ \emph {et~al.}(2018)\citenamefont
  {Hoffmann}, \citenamefont {Appel}, \citenamefont {Rubio},\ and\ \citenamefont
  {Maitra}}]{HARM18}%
  \BibitemOpen
  \bibfield  {author} {\bibinfo {author} {\bibfnamefont {N.~M.}\ \bibnamefont
  {Hoffmann}}, \bibinfo {author} {\bibfnamefont {H.}~\bibnamefont {Appel}},
  \bibinfo {author} {\bibfnamefont {A.}~\bibnamefont {Rubio}}, \ and\ \bibinfo
  {author} {\bibfnamefont {N.~T.}\ \bibnamefont {Maitra}},\ }\href@noop {}
  {\bibfield  {journal} {\bibinfo  {journal} {The European Physical Journal B}\
  }\textbf {\bibinfo {volume} {91}},\ \bibinfo {pages} {180} (\bibinfo {year}
  {2018})}\BibitemShut {NoStop}%
\bibitem [{\citenamefont {Lacombe}\ \emph {et~al.}(2019)\citenamefont
  {Lacombe}, \citenamefont {Hoffmann},\ and\ \citenamefont {Maitra}}]{LHM19}%
  \BibitemOpen
  \bibfield  {author} {\bibinfo {author} {\bibfnamefont {L.}~\bibnamefont
  {Lacombe}}, \bibinfo {author} {\bibfnamefont {N.~M.}\ \bibnamefont
  {Hoffmann}}, \ and\ \bibinfo {author} {\bibfnamefont {N.~T.}\ \bibnamefont
  {Maitra}},\ }\href@noop {} {\bibfield  {journal} {\bibinfo  {journal} {Phys.
  Rev. Lett.}\ }\textbf {\bibinfo {volume} {123}},\ \bibinfo {pages} {083201}
  (\bibinfo {year} {2019})}\BibitemShut {NoStop}%
\bibitem [{\citenamefont {Gidopoulos}\ and\ \citenamefont
  {Gross}(2014)}]{GG14}%
  \BibitemOpen
  \bibfield  {author} {\bibinfo {author} {\bibfnamefont {N.~I.}\ \bibnamefont
  {Gidopoulos}}\ and\ \bibinfo {author} {\bibfnamefont {E.~K.~U.}\ \bibnamefont
  {Gross}},\ }\href@noop {} {\bibfield  {journal} {\bibinfo  {journal}
  {Philosophical Transactions of the Royal Society of London A: Mathematical,
  Physical and Engineering Sciences}\ }\textbf {\bibinfo {volume} {372}}
  (\bibinfo {year} {2014})}\BibitemShut {NoStop}%
\bibitem [{\citenamefont {Hunter}(1986)}]{Hunter86}%
  \BibitemOpen
  \bibfield  {author} {\bibinfo {author} {\bibfnamefont {G.}~\bibnamefont
  {Hunter}},\ }\href@noop {} {\bibfield  {journal} {\bibinfo  {journal} {Int.
  J. Quantum Chem.}\ }\textbf {\bibinfo {volume} {29}},\ \bibinfo {pages} {197}
  (\bibinfo {year} {1986})}\BibitemShut {NoStop}%
\bibitem [{\citenamefont {Buijse}\ \emph {et~al.}(1989)\citenamefont {Buijse},
  \citenamefont {Baerends},\ and\ \citenamefont {Snijders}}]{BBS89}%
  \BibitemOpen
  \bibfield  {author} {\bibinfo {author} {\bibfnamefont {M.~A.}\ \bibnamefont
  {Buijse}}, \bibinfo {author} {\bibfnamefont {E.~J.}\ \bibnamefont
  {Baerends}}, \ and\ \bibinfo {author} {\bibfnamefont {J.~G.}\ \bibnamefont
  {Snijders}},\ }\href@noop {} {\bibfield  {journal} {\bibinfo  {journal}
  {Phys. Rev. A}\ }\textbf {\bibinfo {volume} {40}},\ \bibinfo {pages} {4190}
  (\bibinfo {year} {1989})}\BibitemShut {NoStop}%
\bibitem [{\citenamefont {Schild}\ and\ \citenamefont {Gross}(2017)}]{SG17}%
  \BibitemOpen
  \bibfield  {author} {\bibinfo {author} {\bibfnamefont {A.}~\bibnamefont
  {Schild}}\ and\ \bibinfo {author} {\bibfnamefont {E.~K.~U.}\ \bibnamefont
  {Gross}},\ }\href {\doibase 10.1103/PhysRevLett.118.163202} {\bibfield
  {journal} {\bibinfo  {journal} {Phys. Rev. Lett.}\ }\textbf {\bibinfo
  {volume} {118}},\ \bibinfo {pages} {163202} (\bibinfo {year}
  {2017})}\BibitemShut {NoStop}%
\bibitem [{\citenamefont {Gonze}\ \emph {et~al.}(2018)\citenamefont {Gonze},
  \citenamefont {Zhou},\ and\ \citenamefont {Reining}}]{GZR18}%
  \BibitemOpen
  \bibfield  {author} {\bibinfo {author} {\bibfnamefont {X.}~\bibnamefont
  {Gonze}}, \bibinfo {author} {\bibfnamefont {J.~S.}\ \bibnamefont {Zhou}}, \
  and\ \bibinfo {author} {\bibfnamefont {L.}~\bibnamefont {Reining}},\
  }\href@noop {} {\bibfield  {journal} {\bibinfo  {journal} {The European
  Physical Journal B}\ }\textbf {\bibinfo {volume} {91}},\ \bibinfo {pages}
  {224} (\bibinfo {year} {2018})}\BibitemShut {NoStop}%
\bibitem [{\citenamefont {Lacombe}\ and\ \citenamefont {Maitra}(2020)}]{LM20}%
  \BibitemOpen
  \bibfield  {author} {\bibinfo {author} {\bibfnamefont {L.}~\bibnamefont
  {Lacombe}}\ and\ \bibinfo {author} {\bibfnamefont {N.~T.}\ \bibnamefont
  {Maitra}},\ }\href@noop {} {\bibfield  {journal} {\bibinfo  {journal} {Phys.
  Rev. Lett.}\ }\textbf {\bibinfo {volume} {124}},\ \bibinfo {pages} {206401}
  (\bibinfo {year} {2020})}\BibitemShut {NoStop}%
\bibitem [{\citenamefont {Requist}\ and\ \citenamefont {Gross}(2019)}]{RG19}%
  \BibitemOpen
  \bibfield  {author} {\bibinfo {author} {\bibfnamefont {R.}~\bibnamefont
  {Requist}}\ and\ \bibinfo {author} {\bibfnamefont {E.~K.~U.}\ \bibnamefont
  {Gross}},\ }\href@noop {} {\enquote {\bibinfo {title} {Fock space embedding
  theory for strongly correlated topological phases},}\ } (\bibinfo {year}
  {2019}),\ \Eprint {http://arxiv.org/abs/1909.07933} {arXiv:1909.07933
  [cond-mat.str-el]} \BibitemShut {NoStop}%
\bibitem [{\citenamefont {Suzuki}\ \emph {et~al.}(2014)\citenamefont {Suzuki},
  \citenamefont {Abedi}, \citenamefont {Maitra}, \citenamefont {Yamashita},\
  and\ \citenamefont {Gross}}]{SAMYG14}%
  \BibitemOpen
  \bibfield  {author} {\bibinfo {author} {\bibfnamefont {Y.}~\bibnamefont
  {Suzuki}}, \bibinfo {author} {\bibfnamefont {A.}~\bibnamefont {Abedi}},
  \bibinfo {author} {\bibfnamefont {N.~T.}\ \bibnamefont {Maitra}}, \bibinfo
  {author} {\bibfnamefont {K.}~\bibnamefont {Yamashita}}, \ and\ \bibinfo
  {author} {\bibfnamefont {E.~K.~U.}\ \bibnamefont {Gross}},\ }\href@noop {}
  {\bibfield  {journal} {\bibinfo  {journal} {Phys. Rev. A}\ }\textbf {\bibinfo
  {volume} {89}},\ \bibinfo {pages} {040501(R)} (\bibinfo {year}
  {2014})}\BibitemShut {NoStop}%
\bibitem [{\citenamefont {Abedi}\ \emph {et~al.}(2018)\citenamefont {Abedi},
  \citenamefont {Khosravi},\ and\ \citenamefont {Tokatly}}]{AKT18}%
  \BibitemOpen
  \bibfield  {author} {\bibinfo {author} {\bibfnamefont {A.}~\bibnamefont
  {Abedi}}, \bibinfo {author} {\bibfnamefont {E.}~\bibnamefont {Khosravi}}, \
  and\ \bibinfo {author} {\bibfnamefont {I.~V.}\ \bibnamefont {Tokatly}},\
  }\href@noop {} {\bibfield  {journal} {\bibinfo  {journal} {The European
  Physical Journal B}\ }\textbf {\bibinfo {volume} {91}},\ \bibinfo {pages}
  {194} (\bibinfo {year} {2018})}\BibitemShut {NoStop}%
\bibitem [{\citenamefont {Ruggenthaler}\ \emph {et~al.}(2014)\citenamefont
  {Ruggenthaler}, \citenamefont {Flick}, \citenamefont {Pellegrini},
  \citenamefont {Appel}, \citenamefont {Tokatly},\ and\ \citenamefont
  {Rubio}}]{RFPATR14}%
  \BibitemOpen
  \bibfield  {author} {\bibinfo {author} {\bibfnamefont {M.}~\bibnamefont
  {Ruggenthaler}}, \bibinfo {author} {\bibfnamefont {J.}~\bibnamefont {Flick}},
  \bibinfo {author} {\bibfnamefont {C.}~\bibnamefont {Pellegrini}}, \bibinfo
  {author} {\bibfnamefont {H.}~\bibnamefont {Appel}}, \bibinfo {author}
  {\bibfnamefont {I.~V.}\ \bibnamefont {Tokatly}}, \ and\ \bibinfo {author}
  {\bibfnamefont {A.}~\bibnamefont {Rubio}},\ }\href@noop {} {\bibfield
  {journal} {\bibinfo  {journal} {Physical Review A}\ }\textbf {\bibinfo
  {volume} {90}},\ \bibinfo {pages} {012508} (\bibinfo {year}
  {2014})}\BibitemShut {NoStop}%
\bibitem [{\citenamefont {Tokatly}(2013)}]{T13}%
  \BibitemOpen
  \bibfield  {author} {\bibinfo {author} {\bibfnamefont {I.~V.}\ \bibnamefont
  {Tokatly}},\ }\href@noop {} {\bibfield  {journal} {\bibinfo  {journal} {Phys.
  Rev. Lett.}\ }\textbf {\bibinfo {volume} {110}},\ \bibinfo {pages} {233001}
  (\bibinfo {year} {2013})}\BibitemShut {NoStop}%
\bibitem [{\citenamefont {Mandal}\ \emph {et~al.}(2020)\citenamefont {Mandal},
  \citenamefont {Montillo~Vega},\ and\ \citenamefont {Huo}}]{MVH20}%
  \BibitemOpen
  \bibfield  {author} {\bibinfo {author} {\bibfnamefont {A.}~\bibnamefont
  {Mandal}}, \bibinfo {author} {\bibfnamefont {S.}~\bibnamefont
  {Montillo~Vega}}, \ and\ \bibinfo {author} {\bibfnamefont {P.}~\bibnamefont
  {Huo}},\ }\href {\doibase 10.1021/acs.jpclett.0c02399} {\bibfield  {journal}
  {\bibinfo  {journal} {The Journal of Physical Chemistry Letters}\ }\textbf
  {\bibinfo {volume} {0}},\ \bibinfo {pages} {null} (\bibinfo {year}
  {2020})}\BibitemShut {NoStop}%
\bibitem [{\citenamefont {Hoffmann}\ \emph {et~al.}(2020)\citenamefont
  {Hoffmann}, \citenamefont {Lacombe}, \citenamefont {Rubio},\ and\
  \citenamefont {Maitra}}]{HLRM20}%
  \BibitemOpen
  \bibfield  {author} {\bibinfo {author} {\bibfnamefont {N.~M.}\ \bibnamefont
  {Hoffmann}}, \bibinfo {author} {\bibfnamefont {L.}~\bibnamefont {Lacombe}},
  \bibinfo {author} {\bibfnamefont {A.}~\bibnamefont {Rubio}}, \ and\ \bibinfo
  {author} {\bibfnamefont {N.~T.}\ \bibnamefont {Maitra}},\ }\href@noop {}
  {\bibfield  {journal} {\bibinfo  {journal} {The Journal of Chemical Physics}\
  }\textbf {\bibinfo {volume} {153}},\ \bibinfo {pages} {104103} (\bibinfo
  {year} {2020})}\BibitemShut {NoStop}%
\bibitem [{\citenamefont {Sch\"afer}\ \emph {et~al.}(2020)\citenamefont
  {Sch\"afer}, \citenamefont {Ruggenthaler}, \citenamefont {Rokaj},\ and\
  \citenamefont {Rubio}}]{SRRR20}%
  \BibitemOpen
  \bibfield  {author} {\bibinfo {author} {\bibfnamefont {C.}~\bibnamefont
  {Sch\"afer}}, \bibinfo {author} {\bibfnamefont {M.}~\bibnamefont
  {Ruggenthaler}}, \bibinfo {author} {\bibfnamefont {V.}~\bibnamefont {Rokaj}},
  \ and\ \bibinfo {author} {\bibfnamefont {A.}~\bibnamefont {Rubio}},\ }\href
  {\doibase 10.1021/acsphotonics.9b01649} {\bibfield  {journal} {\bibinfo
  {journal} {ACS Photonics}\ }\textbf {\bibinfo {volume} {7}},\ \bibinfo
  {pages} {975} (\bibinfo {year} {2020})}\BibitemShut {NoStop}%
\bibitem [{\citenamefont {Rokaj}\ \emph {et~al.}(2018)\citenamefont {Rokaj},
  \citenamefont {Welakuh}, \citenamefont {Ruggenthaler},\ and\ \citenamefont
  {Rubio}}]{RWRR18}%
  \BibitemOpen
  \bibfield  {author} {\bibinfo {author} {\bibfnamefont {V.}~\bibnamefont
  {Rokaj}}, \bibinfo {author} {\bibfnamefont {D.~M.}\ \bibnamefont {Welakuh}},
  \bibinfo {author} {\bibfnamefont {M.}~\bibnamefont {Ruggenthaler}}, \ and\
  \bibinfo {author} {\bibfnamefont {A.}~\bibnamefont {Rubio}},\ }\href@noop {}
  {\bibfield  {journal} {\bibinfo  {journal} {Journal of Physics B: Atomic,
  Molecular and Optical Physics}\ }\textbf {\bibinfo {volume} {51}},\ \bibinfo
  {pages} {034005} (\bibinfo {year} {2018})}\BibitemShut {NoStop}%
\bibitem [{\citenamefont {Alonso}\ \emph {et~al.}(2013)\citenamefont {Alonso},
  \citenamefont {Clemente-Gallardo}, \citenamefont {Echenique-Robba},\ and\
  \citenamefont {Jover-Galtier}}]{ACEJ13}%
  \BibitemOpen
  \bibfield  {author} {\bibinfo {author} {\bibfnamefont {J.~L.}\ \bibnamefont
  {Alonso}}, \bibinfo {author} {\bibfnamefont {J.}~\bibnamefont
  {Clemente-Gallardo}}, \bibinfo {author} {\bibfnamefont {P.}~\bibnamefont
  {Echenique-Robba}}, \ and\ \bibinfo {author} {\bibfnamefont {J.~A.}\
  \bibnamefont {Jover-Galtier}},\ }\href@noop {} {\bibfield  {journal}
  {\bibinfo  {journal} {The Journal of Chemical Physics}\ }\textbf {\bibinfo
  {volume} {139}},\ \bibinfo {pages} {087101} (\bibinfo {year}
  {2013})}\BibitemShut {NoStop}%
\bibitem [{\citenamefont {Abedi}\ \emph
  {et~al.}(2013{\natexlab{a}})\citenamefont {Abedi}, \citenamefont {Maitra},\
  and\ \citenamefont {Gross}}]{AMG13}%
  \BibitemOpen
  \bibfield  {author} {\bibinfo {author} {\bibfnamefont {A.}~\bibnamefont
  {Abedi}}, \bibinfo {author} {\bibfnamefont {N.~T.}\ \bibnamefont {Maitra}}, \
  and\ \bibinfo {author} {\bibfnamefont {E.~K.~U.}\ \bibnamefont {Gross}},\
  }\href@noop {} {\bibfield  {journal} {\bibinfo  {journal} {The Journal of
  Chemical Physics}\ }\textbf {\bibinfo {volume} {139}},\ \bibinfo {pages}
  {087102} (\bibinfo {year} {2013}{\natexlab{a}})}\BibitemShut {NoStop}%
\bibitem [{\citenamefont {Khosravi}\ \emph {et~al.}(2015)\citenamefont
  {Khosravi}, \citenamefont {Abedi},\ and\ \citenamefont {Maitra}}]{KAM15}%
  \BibitemOpen
  \bibfield  {author} {\bibinfo {author} {\bibfnamefont {E.}~\bibnamefont
  {Khosravi}}, \bibinfo {author} {\bibfnamefont {A.}~\bibnamefont {Abedi}}, \
  and\ \bibinfo {author} {\bibfnamefont {N.~T.}\ \bibnamefont {Maitra}},\
  }\href@noop {} {\bibfield  {journal} {\bibinfo  {journal} {Phys. Rev. Lett.}\
  }\textbf {\bibinfo {volume} {115}},\ \bibinfo {pages} {263002} (\bibinfo
  {year} {2015})}\BibitemShut {NoStop}%
\bibitem [{\citenamefont {Galego}\ \emph {et~al.}(2015)\citenamefont {Galego},
  \citenamefont {Garcia-Vidal},\ and\ \citenamefont {Feist}}]{GGF15}%
  \BibitemOpen
  \bibfield  {author} {\bibinfo {author} {\bibfnamefont {J.}~\bibnamefont
  {Galego}}, \bibinfo {author} {\bibfnamefont {F.~J.}\ \bibnamefont
  {Garcia-Vidal}}, \ and\ \bibinfo {author} {\bibfnamefont {J.}~\bibnamefont
  {Feist}},\ }\href {\doibase 10.1103/PhysRevX.5.041022} {\bibfield  {journal}
  {\bibinfo  {journal} {Phys. Rev. X}\ }\textbf {\bibinfo {volume} {5}},\
  \bibinfo {pages} {041022} (\bibinfo {year} {2015})}\BibitemShut {NoStop}%
\bibitem [{\citenamefont {Galego}\ \emph {et~al.}(2016)\citenamefont {Galego},
  \citenamefont {Garcia-Vidal},\ and\ \citenamefont {Feist}}]{GGF16}%
  \BibitemOpen
  \bibfield  {author} {\bibinfo {author} {\bibfnamefont {J.}~\bibnamefont
  {Galego}}, \bibinfo {author} {\bibfnamefont {F.~J.}\ \bibnamefont
  {Garcia-Vidal}}, \ and\ \bibinfo {author} {\bibfnamefont {J.}~\bibnamefont
  {Feist}},\ }\href {http://dx.doi.org/10.1038/ncomms13841} {\bibfield
  {journal} {\bibinfo  {journal} {Nature Communications}\ }\textbf {\bibinfo
  {volume} {7}},\ \bibinfo {pages} {13841 EP } (\bibinfo {year}
  {2016})}\BibitemShut {NoStop}%
\bibitem [{\citenamefont {Agostini}\ \emph {et~al.}(2016)\citenamefont
  {Agostini}, \citenamefont {Min}, \citenamefont {Abedi},\ and\ \citenamefont
  {Gross}}]{AMAG16}%
  \BibitemOpen
  \bibfield  {author} {\bibinfo {author} {\bibfnamefont {F.}~\bibnamefont
  {Agostini}}, \bibinfo {author} {\bibfnamefont {S.~K.}\ \bibnamefont {Min}},
  \bibinfo {author} {\bibfnamefont {A.}~\bibnamefont {Abedi}}, \ and\ \bibinfo
  {author} {\bibfnamefont {E.~K.~U.}\ \bibnamefont {Gross}},\ }\href@noop {}
  {\bibfield  {journal} {\bibinfo  {journal} {Journal of Chemical Theory and
  Computation}\ }\textbf {\bibinfo {volume} {12}},\ \bibinfo {pages} {2127}
  (\bibinfo {year} {2016})}\BibitemShut {NoStop}%
\bibitem [{\citenamefont {Agostini}\ \emph {et~al.}(2015)\citenamefont
  {Agostini}, \citenamefont {Abedi}, \citenamefont {Suzuki}, \citenamefont
  {Min}, \citenamefont {Maitra},\ and\ \citenamefont {Gross}}]{AASMMG15}%
  \BibitemOpen
  \bibfield  {author} {\bibinfo {author} {\bibfnamefont {F.}~\bibnamefont
  {Agostini}}, \bibinfo {author} {\bibfnamefont {A.}~\bibnamefont {Abedi}},
  \bibinfo {author} {\bibfnamefont {Y.}~\bibnamefont {Suzuki}}, \bibinfo
  {author} {\bibfnamefont {S.~K.}\ \bibnamefont {Min}}, \bibinfo {author}
  {\bibfnamefont {N.~T.}\ \bibnamefont {Maitra}}, \ and\ \bibinfo {author}
  {\bibfnamefont {E.~K.~U.}\ \bibnamefont {Gross}},\ }\href@noop {} {\bibfield
  {journal} {\bibinfo  {journal} {J. Chem. Phys.}\ }\textbf {\bibinfo {volume}
  {142}},\ \bibinfo {pages} {084303} (\bibinfo {year} {2015})}\BibitemShut
  {NoStop}%
\bibitem [{\citenamefont {Gossel}\ \emph {et~al.}(2018)\citenamefont {Gossel},
  \citenamefont {Agostini},\ and\ \citenamefont {Maitra}}]{GAM18}%
  \BibitemOpen
  \bibfield  {author} {\bibinfo {author} {\bibfnamefont {G.~H.}\ \bibnamefont
  {Gossel}}, \bibinfo {author} {\bibfnamefont {F.}~\bibnamefont {Agostini}}, \
  and\ \bibinfo {author} {\bibfnamefont {N.~T.}\ \bibnamefont {Maitra}},\
  }\href@noop {} {\bibfield  {journal} {\bibinfo  {journal} {Journal of
  Chemical Theory and Computation}\ }\textbf {\bibinfo {volume} {14}},\
  \bibinfo {pages} {4513} (\bibinfo {year} {2018})}\BibitemShut {NoStop}%
\bibitem [{\citenamefont {Agostini}\ and\ \citenamefont
  {Curchod}(2019)}]{AC19}%
  \BibitemOpen
  \bibfield  {author} {\bibinfo {author} {\bibfnamefont {F.}~\bibnamefont
  {Agostini}}\ and\ \bibinfo {author} {\bibfnamefont {B.~F.~E.}\ \bibnamefont
  {Curchod}},\ }\href@noop {} {\bibfield  {journal} {\bibinfo  {journal} {Wiley
  Interdisciplinary Reviews: Computational Molecular Science}\ }\textbf
  {\bibinfo {volume} {0}},\ \bibinfo {pages} {e1417} (\bibinfo {year}
  {2019})}\BibitemShut {NoStop}%
\bibitem [{\citenamefont {Abedi}\ \emph
  {et~al.}(2013{\natexlab{b}})\citenamefont {Abedi}, \citenamefont {Agostini},
  \citenamefont {Suzuki},\ and\ \citenamefont {Gross}}]{AASG13}%
  \BibitemOpen
  \bibfield  {author} {\bibinfo {author} {\bibfnamefont {A.}~\bibnamefont
  {Abedi}}, \bibinfo {author} {\bibfnamefont {F.}~\bibnamefont {Agostini}},
  \bibinfo {author} {\bibfnamefont {Y.}~\bibnamefont {Suzuki}}, \ and\ \bibinfo
  {author} {\bibfnamefont {E.~K.~U.}\ \bibnamefont {Gross}},\ }\href@noop {}
  {\bibfield  {journal} {\bibinfo  {journal} {Phys. Rev. Lett.}\ }\textbf
  {\bibinfo {volume} {110}},\ \bibinfo {pages} {263001} (\bibinfo {year}
  {2013}{\natexlab{b}})}\BibitemShut {NoStop}%
\bibitem [{\citenamefont {Gossel}\ \emph {et~al.}(2019)\citenamefont {Gossel},
  \citenamefont {Lacombe},\ and\ \citenamefont {Maitra}}]{GLM19}%
  \BibitemOpen
  \bibfield  {author} {\bibinfo {author} {\bibfnamefont {G.~H.}\ \bibnamefont
  {Gossel}}, \bibinfo {author} {\bibfnamefont {L.}~\bibnamefont {Lacombe}}, \
  and\ \bibinfo {author} {\bibfnamefont {N.~T.}\ \bibnamefont {Maitra}},\
  }\href@noop {} {\bibfield  {journal} {\bibinfo  {journal} {The Journal of
  Chemical Physics}\ }\textbf {\bibinfo {volume} {150}},\ \bibinfo {pages}
  {154112} (\bibinfo {year} {2019})}\BibitemShut {NoStop}%
\bibitem [{\citenamefont {Lorin}(2020)}]{Lorin20}%
  \BibitemOpen
  \bibfield  {author} {\bibinfo {author} {\bibfnamefont {E.}~\bibnamefont
  {Lorin}},\ }\href@noop {} {\enquote {\bibinfo {title} {Numerical analysis of
  the exact factorization of molecular time-dependent schr\"odinger
  wavefunctions},}\ } (\bibinfo {year} {2020})\BibitemShut {NoStop}%
\bibitem [{\citenamefont {Min}\ \emph {et~al.}(2015)\citenamefont {Min},
  \citenamefont {Agostini},\ and\ \citenamefont {Gross}}]{MAG15}%
  \BibitemOpen
  \bibfield  {author} {\bibinfo {author} {\bibfnamefont {S.~K.}\ \bibnamefont
  {Min}}, \bibinfo {author} {\bibfnamefont {F.}~\bibnamefont {Agostini}}, \
  and\ \bibinfo {author} {\bibfnamefont {E.~K.~U.}\ \bibnamefont {Gross}},\
  }\href@noop {} {\bibfield  {journal} {\bibinfo  {journal} {Phys. Rev. Lett.}\
  }\textbf {\bibinfo {volume} {115}},\ \bibinfo {pages} {073001} (\bibinfo
  {year} {2015})}\BibitemShut {NoStop}%
\bibitem [{\citenamefont {Min}\ \emph {et~al.}(2017)\citenamefont {Min},
  \citenamefont {Agostini}, \citenamefont {Tavernelli},\ and\ \citenamefont
  {Gross}}]{MATG17}%
  \BibitemOpen
  \bibfield  {author} {\bibinfo {author} {\bibfnamefont {S.~K.}\ \bibnamefont
  {Min}}, \bibinfo {author} {\bibfnamefont {F.}~\bibnamefont {Agostini}},
  \bibinfo {author} {\bibfnamefont {I.}~\bibnamefont {Tavernelli}}, \ and\
  \bibinfo {author} {\bibfnamefont {E.~K.~U.}\ \bibnamefont {Gross}},\
  }\href@noop {} {\bibfield  {journal} {\bibinfo  {journal} {The Journal of
  Physical Chemistry Letters}\ }\textbf {\bibinfo {volume} {8}},\ \bibinfo
  {pages} {3048} (\bibinfo {year} {2017})}\BibitemShut {NoStop}%
\bibitem [{\citenamefont {Ha}\ \emph {et~al.}(2018)\citenamefont {Ha},
  \citenamefont {Lee},\ and\ \citenamefont {Min}}]{HLM18}%
  \BibitemOpen
  \bibfield  {author} {\bibinfo {author} {\bibfnamefont {J.-K.}\ \bibnamefont
  {Ha}}, \bibinfo {author} {\bibfnamefont {I.~S.}\ \bibnamefont {Lee}}, \ and\
  \bibinfo {author} {\bibfnamefont {S.~K.}\ \bibnamefont {Min}},\ }\href@noop
  {} {\bibfield  {journal} {\bibinfo  {journal} {The Journal of Physical
  Chemistry Letters}\ }\textbf {\bibinfo {volume} {9}},\ \bibinfo {pages}
  {1097} (\bibinfo {year} {2018})}\BibitemShut {NoStop}%
\bibitem [{\citenamefont {Filatov}\ \emph
  {et~al.}(2019{\natexlab{a}})\citenamefont {Filatov}, \citenamefont {Min},\
  and\ \citenamefont {Kim}}]{FMK19}%
  \BibitemOpen
  \bibfield  {author} {\bibinfo {author} {\bibfnamefont {M.}~\bibnamefont
  {Filatov}}, \bibinfo {author} {\bibfnamefont {S.~K.}\ \bibnamefont {Min}}, \
  and\ \bibinfo {author} {\bibfnamefont {K.~S.}\ \bibnamefont {Kim}},\ }\href
  {\doibase 10.1080/00268976.2018.1519200} {\bibfield  {journal} {\bibinfo
  {journal} {Molecular Physics}\ }\textbf {\bibinfo {volume} {117}},\ \bibinfo
  {pages} {1128} (\bibinfo {year} {2019}{\natexlab{a}})}\BibitemShut {NoStop}%
\bibitem [{\citenamefont {Filatov}\ \emph
  {et~al.}(2019{\natexlab{b}})\citenamefont {Filatov}, \citenamefont {Paolino},
  \citenamefont {Min},\ and\ \citenamefont {Choi}}]{FPMC19}%
  \BibitemOpen
  \bibfield  {author} {\bibinfo {author} {\bibfnamefont {M.}~\bibnamefont
  {Filatov}}, \bibinfo {author} {\bibfnamefont {M.}~\bibnamefont {Paolino}},
  \bibinfo {author} {\bibfnamefont {S.~K.}\ \bibnamefont {Min}}, \ and\
  \bibinfo {author} {\bibfnamefont {C.~H.}\ \bibnamefont {Choi}},\ }\href@noop
  {} {\bibfield  {journal} {\bibinfo  {journal} {Chem. Commun.}\ }\textbf
  {\bibinfo {volume} {55}},\ \bibinfo {pages} {5247} (\bibinfo {year}
  {2019}{\natexlab{b}})}\BibitemShut {NoStop}%
\bibitem [{\citenamefont {Filatov}\ \emph {et~al.}(2018)\citenamefont
  {Filatov}, \citenamefont {Paolino}, \citenamefont {Min},\ and\ \citenamefont
  {Kim}}]{FPMK18}%
  \BibitemOpen
  \bibfield  {author} {\bibinfo {author} {\bibfnamefont {M.}~\bibnamefont
  {Filatov}}, \bibinfo {author} {\bibfnamefont {M.}~\bibnamefont {Paolino}},
  \bibinfo {author} {\bibfnamefont {S.~K.}\ \bibnamefont {Min}}, \ and\
  \bibinfo {author} {\bibfnamefont {K.~S.}\ \bibnamefont {Kim}},\ }\href
  {\doibase 10.1021/acs.jpclett.8b02268} {\bibfield  {journal} {\bibinfo
  {journal} {The Journal of Physical Chemistry Letters}\ }\textbf {\bibinfo
  {volume} {9}},\ \bibinfo {pages} {4995} (\bibinfo {year} {2018})}\BibitemShut
  {NoStop}%
\bibitem [{\citenamefont {Shin}\ and\ \citenamefont {Metiu}(1995)}]{SM95}%
  \BibitemOpen
  \bibfield  {author} {\bibinfo {author} {\bibfnamefont {S.}~\bibnamefont
  {Shin}}\ and\ \bibinfo {author} {\bibfnamefont {H.}~\bibnamefont {Metiu}},\
  }\href@noop {} {\bibfield  {journal} {\bibinfo  {journal} {The Journal of
  Chemical Physics}\ }\textbf {\bibinfo {volume} {102}},\ \bibinfo {pages}
  {9285} (\bibinfo {year} {1995})}\BibitemShut {NoStop}%
\bibitem [{\citenamefont {Fang}\ and\ \citenamefont
  {Hammes-Schiffer}(1997{\natexlab{a}})}]{FH97}%
  \BibitemOpen
  \bibfield  {author} {\bibinfo {author} {\bibfnamefont {J.-Y.}\ \bibnamefont
  {Fang}}\ and\ \bibinfo {author} {\bibfnamefont {S.}~\bibnamefont
  {Hammes-Schiffer}},\ }\href@noop {} {\bibfield  {journal} {\bibinfo
  {journal} {The Journal of Chemical Physics}\ }\textbf {\bibinfo {volume}
  {106}},\ \bibinfo {pages} {8442} (\bibinfo {year}
  {1997}{\natexlab{a}})}\BibitemShut {NoStop}%
\bibitem [{\citenamefont {Fang}\ and\ \citenamefont
  {Hammes-Schiffer}(1997{\natexlab{b}})}]{FH97b}%
  \BibitemOpen
  \bibfield  {author} {\bibinfo {author} {\bibfnamefont {J.-Y.}\ \bibnamefont
  {Fang}}\ and\ \bibinfo {author} {\bibfnamefont {S.}~\bibnamefont
  {Hammes-Schiffer}},\ }\href@noop {} {\bibfield  {journal} {\bibinfo
  {journal} {The Journal of Chemical Physics}\ }\textbf {\bibinfo {volume}
  {107}},\ \bibinfo {pages} {5727} (\bibinfo {year}
  {1997}{\natexlab{b}})}\BibitemShut {NoStop}%
\bibitem [{\citenamefont {Heller}(1976)}]{H76}%
  \BibitemOpen
  \bibfield  {author} {\bibinfo {author} {\bibfnamefont {E.~J.}\ \bibnamefont
  {Heller}},\ }\href {\doibase 10.1063/1.433238} {\bibfield  {journal}
  {\bibinfo  {journal} {The Journal of Chemical Physics}\ }\textbf {\bibinfo
  {volume} {65}},\ \bibinfo {pages} {1289} (\bibinfo {year}
  {1976})}\BibitemShut {NoStop}%
\bibitem [{\citenamefont {Curchod}\ \emph {et~al.}(2016)\citenamefont
  {Curchod}, \citenamefont {Agostini},\ and\ \citenamefont {Gross}}]{CAG16}%
  \BibitemOpen
  \bibfield  {author} {\bibinfo {author} {\bibfnamefont {B.~F.~E.}\
  \bibnamefont {Curchod}}, \bibinfo {author} {\bibfnamefont {F.}~\bibnamefont
  {Agostini}}, \ and\ \bibinfo {author} {\bibfnamefont {E.~K.~U.}\ \bibnamefont
  {Gross}},\ }\href@noop {} {\bibfield  {journal} {\bibinfo  {journal} {J.
  Chem. Phys.}\ }\textbf {\bibinfo {volume} {145}},\ \bibinfo {pages} {034103}
  (\bibinfo {year} {2016})}\BibitemShut {NoStop}%
\bibitem [{\citenamefont {Flick}\ \emph {et~al.}(2017)\citenamefont {Flick},
  \citenamefont {Appel}, \citenamefont {Ruggenthaler},\ and\ \citenamefont
  {Rubio}}]{FARR17}%
  \BibitemOpen
  \bibfield  {author} {\bibinfo {author} {\bibfnamefont {J.}~\bibnamefont
  {Flick}}, \bibinfo {author} {\bibfnamefont {H.}~\bibnamefont {Appel}},
  \bibinfo {author} {\bibfnamefont {M.}~\bibnamefont {Ruggenthaler}}, \ and\
  \bibinfo {author} {\bibfnamefont {A.}~\bibnamefont {Rubio}},\ }\href@noop {}
  {\bibfield  {journal} {\bibinfo  {journal} {Journal of Chemical Theory and
  Computation}\ }\textbf {\bibinfo {volume} {13}},\ \bibinfo {pages} {1616}
  (\bibinfo {year} {2017})}\BibitemShut {NoStop}%
\bibitem [{\citenamefont {Curchod}\ \emph {et~al.}(2018)\citenamefont
  {Curchod}, \citenamefont {Agostini},\ and\ \citenamefont
  {Tavernelli}}]{CAT18}%
  \BibitemOpen
  \bibfield  {author} {\bibinfo {author} {\bibfnamefont {B.}~\bibnamefont
  {Curchod}}, \bibinfo {author} {\bibfnamefont {F.}~\bibnamefont {Agostini}}, \
  and\ \bibinfo {author} {\bibfnamefont {I.}~\bibnamefont {Tavernelli}},\
  }\href@noop {} {\bibfield  {journal} {\bibinfo  {journal} {The European
  Physical Journal B}\ }\textbf {\bibinfo {volume} {91}},\ \bibinfo {pages}
  {168} (\bibinfo {year} {2018})}\BibitemShut {NoStop}%
\bibitem [{\citenamefont {Groenhof}\ \emph {et~al.}(2019)\citenamefont
  {Groenhof}, \citenamefont {Climent}, \citenamefont {Feist}, \citenamefont
  {Morozov},\ and\ \citenamefont {Toppari}}]{GCFMT2019}%
  \BibitemOpen
  \bibfield  {author} {\bibinfo {author} {\bibfnamefont {G.}~\bibnamefont
  {Groenhof}}, \bibinfo {author} {\bibfnamefont {C.}~\bibnamefont {Climent}},
  \bibinfo {author} {\bibfnamefont {J.}~\bibnamefont {Feist}}, \bibinfo
  {author} {\bibfnamefont {D.}~\bibnamefont {Morozov}}, \ and\ \bibinfo
  {author} {\bibfnamefont {J.~J.}\ \bibnamefont {Toppari}},\ }\href {\doibase
  10.1021/acs.jpclett.9b02192} {\bibfield  {journal} {\bibinfo  {journal} {The
  Journal of Physical Chemistry Letters}\ }\textbf {\bibinfo {volume} {10}},\
  \bibinfo {pages} {5476} (\bibinfo {year} {2019})}\BibitemShut {NoStop}%
\end{thebibliography}%

\end{document}